\documentclass[journal]{IEEEtran}

\usepackage{cite}
\usepackage{hyperref}
\ifCLASSINFOpdf
\else
\fi
\usepackage{amsmath}
\usepackage{booktabs}

\ifCLASSOPTIONcompsoc
 \usepackage[caption=false,font=normalsize,labelfont=sf,textfont=sf]{subfig}
\else
 \usepackage[caption=false,font=footnotesize]{subfig}
\fi

\usepackage{url}

\usepackage{amssymb}
\usepackage{graphicx}
\usepackage{algorithm}
\usepackage{algpseudocode}

\begin{document}

\title{Deep Bilinear Koopman Model for Real-Time Vehicle Control in Frenet Frame}

\author{Mohammad Abtahi,~\IEEEmembership{Member,~IEEE,}, Farhang Motallebi Araghi, Navid Mojahed, 
        Shima Nazari,~\IEEEmembership{Member,~IEEE,}% <-this % stops a space
\thanks{ 

The authors are with the Department of Mechanical and Aerospace Engineering, University of California Davis, USA (e-mail: sabtahi@ucdavis.edu; fmotallebi@ucdavis.edu; nmojahed@ucdavis.edu;  snazari@ucdavis.edu)}}

\maketitle

\begin{abstract}
Accurate modeling and control of autonomous vehicles remain a fundamental challenge due to the nonlinear and coupled nature of vehicle dynamics. While Koopman operator theory offers a framework for deploying powerful linear control techniques, learning a finite-dimensional invariant subspace for high-fidelity modeling continues to be an open problem.
This paper presents a deep Koopman approach for modeling and control of vehicle dynamics within the curvilinear Frenet frame. The proposed framework uses a deep neural network architecture to simultaneously learn the Koopman operator and its associated invariant subspace from the data.  Input–state bilinear interactions are captured by the algorithm while preserving convexity, which makes it suitable for real-time model predictive control (MPC) application. A multi-step prediction loss is utilized during training to ensure long-horizon prediction capability. To further enhance real-time trajectory tracking performance, the model is integrated with a cumulative error regulator (CER) module, which compensates for model mismatch by mitigating accumulated prediction errors.
Closed-loop performance is evaluated through hardware-in-the-loop (HIL) experiments using a CarSim RT model as the target plant, with real-time validation conducted on a dSPACE SCALEXIO system. The proposed controller achieved significant reductions in tracking error relative to baseline controllers, confirming its suitability for real-time implementation in embedded autonomous vehicle systems.

\end{abstract}


% \fi
\IEEEpeerreviewmaketitle

\section{Introduction}
\IEEEPARstart{A}{utonomous} vehicles have the potential to revolutionize transportation, but ensuring safety remains a critical challenge. Achieving safe and reliable operation under diverse and potentially hazardous conditions requires accurate modeling and robust control strategies \cite{othman2022exploring}.
However, real-time control development is hindered by the nonlinear nature of vehicle dynamics and the inherent uncertainties in real-world driving environments.

Optimization-based control strategies such as Model Predictive Control (MPC) \cite{paden2016survey, maciejowski2002predictive, abtahi2023automatic} and Linear-Quadratic Regulators (LQR) \cite{xing2024control} stand out for enhancing the safety and maneuvering capabilities of autonomous vehicles \cite{marti2025optimization}. In particular, nonlinear MPC offers high-fidelity control performance by directly accounting for vehicle and tire nonlinearities, but it typically results in a non-convex optimization problem. This substantially increases computational demands and may lead to suboptimal outcomes, limiting the applicability of nonlinear MPC in real-time scenarios \cite{attia2014nonlinear}. On the other hand, linear control approaches, such as LQR and Linear MPC, offer computational efficiency and theoretical guarantees of optimality by leveraging simplified models of vehicle dynamics~\cite{jannah2022nonlinear}. These models are typically obtained through Jacobian linearization around equilibrium points or along trajectory operating points, resulting in linear time-invariant (LTI) or linear time-varying (LTV) representations~\cite{nugroho2024vehicle}. While effective for local approximations, such linearized models often fail to capture the full nonlinear behavior of vehicle dynamics, specifically during aggressive steering maneuvers, emergency braking, or operation at stability limits \cite{goh2024beyond}. These limitations motivate the exploration of alternative modeling techniques that balance accuracy and computational efficiency.
\begin{figure}
   \centering
   \includegraphics[width=0.8\linewidth]{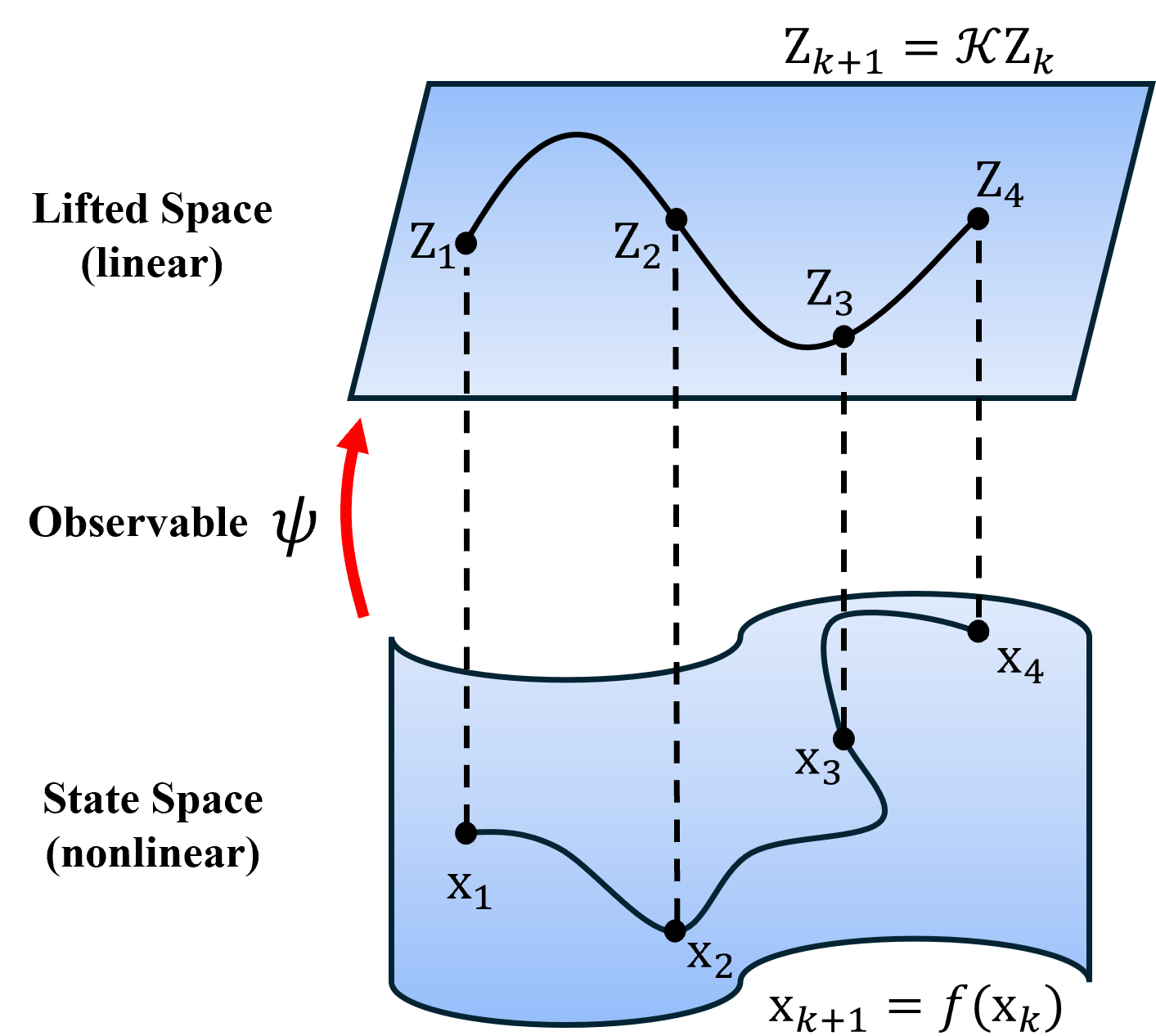} % Directly include image
   \caption{Illustration of Koopman operator lifting.}
   \label{fig:Koopman_Lifting}
\end{figure}

Recent advances in data collection and large-scale datasets have made data-driven modeling effective for learning system behaviors \cite{brunton2022data}. These methods rely directly on observed data rather than first-principles modeling \cite{brunton2021modern}. In particular, end-to-end learning approaches, where raw sensor data are directly mapped to control actions or vehicle states using deep neural networks, have been widely explored in the context of autonomous driving and vehicle dynamics modeling \cite{bojarski2016end}. While such models often achieve high predictive performance, they tend to lack interpretability and structure, making them difficult to analyze or incorporate into safety-critical control systems. Among interpretable alternatives, the Koopman operator theory has gained significant attention for its ability to transform nonlinear dynamics into an infinite-dimensional linear representation in the space of observable functions \cite{koopman1931hamiltonian}. These scalar-valued functions lift the original state space into a higher-dimensional space, allowing nonlinear system behavior to be described using linear structures, as illustrated in Fig.~\ref{fig:Koopman_Lifting}. Such a linear representation allows the direct application of linear control strategies to inherently nonlinear systems, making Koopman operator methods particularly promising for AV modeling and control \cite{korda2018linear}.
Since numerical computation with an infinite-dimensional Koopman operator is impractical, researchers have developed several techniques to approximate a finite-dimensional representation for Koopman dynamics. Dynamic Mode Decomposition (DMD) \cite{kutz2016dynamic, takeishi2017learning}, Extended Dynamic Mode Decomposition (EDMD)\cite{williams2016extending}, and the eigenfunction approach \cite{ brunton2016discovering} have been widely used to approximate Koopman representations of systems with no control inputs. In addition, variants of DMD and EDMD, called DMDc \cite{proctor2016dynamic} and EDMDc \cite{korda2018linear}, have been extensively applied to systems with control inputs.

Despite recent advancements, most Koopman-based control methods assume linear dependence on control inputs, which limits their ability to capture nonlinear control-state coupling and reduces predictive accuracy in general nonlinear systems. This limitation becomes particularly problematic in vehicle dynamics, where, for instance, the steering angle interacts nonlinearly with vehicle speed to generate lateral forces.
On the other hand, embedding nonlinear control inputs within the lifting functions has been shown to introduce causality issues \cite{selby2021physics}, which remains an open challenge in the field. As an alternative, bilinear models provide a favorable trade-off between capturing nonlinear system dynamics and maintaining computational tractability \cite{bruder2021advantages}. However, employing a bilinear predictive model can lead to non-convexities in the resulting optimal control problem. Prior works have addressed this challenge by fixing the coupling terms over the prediction horizon to preserve linearity \cite{yu2022autonomous}.

Several studies have explored Koopman-based methods for vehicle dynamics modeling and control. Cibulka et al. \cite{cibulka2019data} identified velocity dynamics using EDMD with manually selected basis functions. The authors later extended this approach to a predictive controller \cite{cibulka2020model}. Svec et al.\cite{vsvec2021model} proposed an MPC scheme using a predictor identified via EDMD and fourth-order polynomial basis functions, demonstrating improved predictive performance over an MPC controller with a linearized model. Kim et al.\cite{kim2022data} applied EDMD to obtain a Koopman model of vehicle dynamics for an LQR lane-keeping controller, achieving improved accuracy over traditional linear models. More recently, Yu et al. \cite{yu2022autonomous} proposed a Koopman MPC framework for motion planning and control using polynomial observables. 

\begin{figure}
   \centering
   \includegraphics[width=0.8\linewidth]{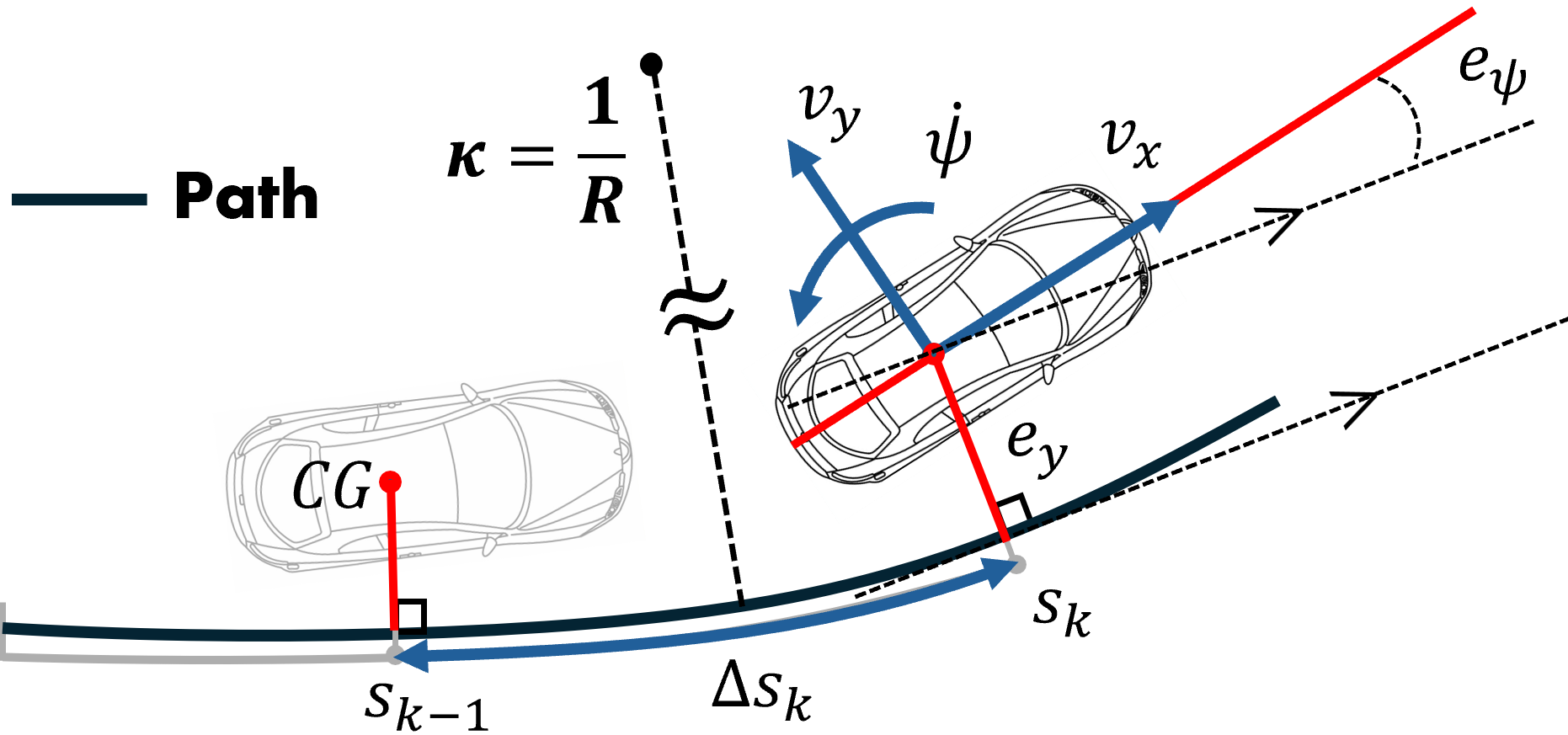} % Directly include image
   \caption{Vehicle states in the curvilinear Frenet frame: longitudinal velocity \(v_x\), lateral velocity \(v_y\), yaw rate \(\dot{\psi}\), progress gain \(\Delta s\), lateral deviation \(e_y\), heading error \(e_\psi\). Road curvature is denoted by \(\kappa\).}
   \label{fig:fernet frame}
\end{figure}

Despite these advancements, a significant limitation of these approaches is the manual selection of observables, which often relies on trial-and-error procedures to identify lifting functions that adequately span the Koopman-invariant subspace. With the emergence of large-scale AI applications and the high representational capacity of neural networks, recent studies have investigated deep learning approaches to address this issue. These approaches automate the identification of Koopman observables through autoencoder structures \cite{lusch2018deep}. Xiao et al. \cite{xiao2022deep, xiao2023ddk} proposed a neural network model to learn Koopman representations for vehicle dynamics. While effective for tracking velocity profiles, the method lacked a global positioning framework, which is essential for achieving high-level navigation and control objectives. To address this limitation, the approach proposed in our previous work \cite{abtahi2025multi} introduced a deep Koopman framework that models vehicle dynamics within the curvilinear Frenet frame illustrated in Fig. \ref{fig:fernet frame}.
Building on that foundation, the present work extends the model to a bilinear structure. This novel scheme enables the encoding of bilinear interactions between inputs and state-dependent observables while preserving compatibility with predictive control strategies. To further increase control precision, the identified dynamics are augmented with a Cumulative Error Regulator (CER), which compensates for model mismatch during optimization by explicitly propagating and penalizing accumulated prediction errors over the horizon. The prediction capability of the proposed approach is then compared against an LTI model, an EDMD Koopman model, and the previously developed Multi-Step Deep Koopman Network (MDK-Net) structure \cite{abtahi2025multi}. Unlike prior approaches, our method directly incorporates driver inputs, steering wheel angle, throttle, and brake pedal commands as system inputs. This design choice enables capturing the nonlinearities within the steering actuation, powertrain dynamics, driveline, and braking behavior. Finally, hardware-in-the-loop (HIL) simulations are conducted using a dSPACE SCALEXIO platform to ensure the real-time applicability of the proposed controller. The controller's real-time performance is then benchmarked against baseline controllers using a high-fidelity CarSim model.

The key contributions of this work are outlined below:
\begin{itemize}\setlength{\itemsep}{4pt}
    \item A Multi-Step Deep Bilinear Koopman Network (MDBK-Net) is developed, which jointly learns the Koopman-invariant subspace via an encoder network and the bilinear Koopman operator matrices.
    \item To enhance model fidelity in capturing the nonlinear dynamics of the steering, powertrain, driveline, and braking subsystems, the driver commands of steering wheel angle, throttle, and brake pedal are directly incorporated as system inputs. Additionally, Frenet frame modeling is achieved by integrating road curvature as an exogenous input, which facilitates the coupled vehicle lateral and longitudinal control.
    \item The identified lifted dynamics are augmented with the CER module within the MPC formulation. This module is designed to mitigate cumulative tracking errors arising from the Koopman approximation, thereby further enhancing the controller’s tracking performance and safety.
    \item The real-time performance of the proposed controller is validated through HIL testing using a dSPACE SCALEXIO system and a high-fidelity CarSim Real-Time (RT) vehicle model, demonstrating the controller’s computational efficiency for real-time implementation.
\end{itemize}

This work highlights a strong use case for leveraging AI, particularly deep learning, in addressing longstanding challenges in model identification and real-time control of complex nonlinear systems. The remainder of the paper is organized as follows. Section \ref{sec:Preliminaries } presents the necessary preliminaries to Koopman operator theory. The structure of the proposed MDBK-Net, its augmentation with CER, and the developed MPC framework are all discussed in Section \ref{sec:Methodology}. Section \ref{sec:Results And Experimental Setup} describes the experimental setup and presents the evaluation results, and conclusions are drawn in Section \ref{sec:Conclusion}

\section{Preliminaries } \label{sec:Preliminaries }
This section introduces the Koopman operator theory and its application to nonlinear system modeling, forming the theoretical foundation for the proposed deep bilinear Koopman control framework. The Koopman operator, originally introduced for dynamical systems analysis \cite{koopman1931hamiltonian, mezic2005spectral}, provides a globally linear representation of nonlinear dynamical systems by lifting the states of a system to a higher-dimensional space of observables. Such a linear representation enables the application of well-established linear control techniques to inherently nonlinear systems.

% We begin by outlining the Koopman operator formulation for both autonomous and control-affine systems. Then, we present the finite-dimensional approximation using a linear Koopman model, followed by its bilinear extension to better capture state-control interactions. This bilinear structure maintains optimization tractability while enriching the expressiveness of the model—striking a balance that is critical for accurate prediction and efficient controller design in real-world autonomous vehicle applications.
%In the following sections, we first introduce the vehicle dynamics model implemented in this work, followed by a concise overview of the Koopman operator theory and its identification for vehicle dynamics.

%%%%%%%%%%%%%%%%%%%%%%%%%%%%%%%%%%%%%%%%%%%%%%%%%%%%%%%%%%%%%%%%%%%%%%%%

\subsection{Koopman Operator }
% Initially, Koopman operator theory was introduced to identify an infinite dimensional linear representation for autonomous dynamics \cite{koopman1931hamiltonian}.
Consider a discrete-time nonlinear autonomous system as
\begin{equation}
\mathbf{x}_{k+1} = \boldsymbol{f_a}(\mathbf{x}_k),
\label{eq:autonomousnonlinear_dynamics}
\end{equation}
where $\mathbf{x}_{k} \in \mathbb{R}^n $ is the state vector of the system at time step $k$, and $f_a :\mathbb{R}^n \to \mathbb{R}^n $ is the nonlinear smooth mapping of the dynamics. Then the infinite dimensional Koopman operator $\mathcal{K}_a:\mathcal{H}_a \to \mathcal{H}_a $ provides a linear representation of the nonlinear dynamics in (\ref{eq:autonomousnonlinear_dynamics}) by acting on scalar-valued observable functions $\psi_a(\cdot): \mathbb{R}^n \to \mathbb{R} $ as 
\begin{equation}
    \psi_a(\mathbf{x}_{k+1}) = \psi_a(\boldsymbol{f_a}(\mathbf{x}_{k}))=\mathcal{K}_a \psi_a(\mathbf{x}_{k}),
\label{eq:koopman_Autonomous }
\end{equation}
where each observable function $\psi_a(\cdot)\in \mathcal{H}_a$ is an element of infinite dimensional Hilbert space $\mathcal{H}_a$, which is invariant under the Koopman operator \cite{korda2018linear}.
Although the Koopman operator was originally formulated for autonomous systems \cite{koopman1931hamiltonian}, works by Korda et al. \cite{korda2018linear} and Proctor et al. \cite{proctor2018generalizing} have shown that it can also be extended to input driven systems. Consider a general discrete-time nonlinear system with control input as  
\begin{equation}
\mathbf{x}_{k+1} = \boldsymbol{f}(\mathbf{x}_k, \mathbf{u}_k),
\label{eq:general_nonlinear_dynamics}
\end{equation}
where $\mathbf{u}_k\in\mathbb{R}^m$ is the control input vector at time $k$. The space of admissible control sequences is defined as $\mathcal{L}(\mathcal{U})=\{(u_k)^\infty_{k=0}| u_k\in\mathcal{U}\}$ which was introduced in \cite{korda2018linear}. The extended state-space can then be defined in the space of $\mathbb{R}^n \times \mathcal{L}(\mathcal{U})$, and it is given by 
\begin{equation}
    \label{extended state}
    \chi_k=
    \begin{bmatrix}
         \mathbf{x}_k \\
         \mathbf{\nu}_k
    \end{bmatrix},
\end{equation}
where $\mathbf{\nu}_k = [\mathbf{u}_k,\mathbf{u}_{k+1},...\mathbf{u}_{\infty} ] \subset \mathcal{L}(\mathcal{U})$ is an infinite dimensional sequence of input vector. Under this formulation, the system dynamics (\ref{eq:general_nonlinear_dynamics}) can be described as an autonomous system such that

\begin{equation}
\chi_{k+1} = F(\chi_{k}) := 
\begin{bmatrix} 
\boldsymbol{f}(\mathbf{x}_{k}, \mathbf{\nu}_{k}(0))  \\ 
 \mathcal{S}\mathbf{\nu}_{k} 
\end{bmatrix},
\label{augmented-autnomous control}
\end{equation}
where $\mathcal{S}$ is the left shift operator, i.e. $\mathcal{S}\mathbf{\nu}_{k}=\mathbf{\nu}_{k+1}$, and $\mathbf{\nu}_{k}(0)=\mathbf{u}_k$ is the first element of the control sequence of $\mathbf{\nu}$ at time step $k$. Then the infinite dimensional Koopman operator $\mathcal{K}: \mathcal{H} \to \mathcal{H} $ associated with (\ref{augmented-autnomous control}) can be defined as
\begin{equation}
    \psi(\chi_{k+1}) = \psi(F(\chi_k))= \mathcal{K}\psi(\chi_k),
    \label{eq:augmented_linear_koopman}
\end{equation}
where $\psi(\cdot) \in \mathcal{H}:\mathbb{R}^n \times \mathcal{L}(\mathcal{U}) \to \mathbb{R}$ is a real-valued observable function that belongs to the extended function space of $\mathcal{H}$ invariant under Koopman operator $\mathcal{K}$. Equation (\ref{eq:augmented_linear_koopman}) implies that the Koopman operator is linear in the function space $\mathcal{H}$, even when the dynamic system $\boldsymbol{f}(., .)$  is nonlinear \cite{mauroy2020koopman}. 

% \( \mathcal{H} \) denotes a Hilbert space of scalar-valued observable functions, typically assumed to be \( L^2(\mathbb{R}^n, \mu) \), the space of square-integrable functions with respect to a measure \( \mu \) on the state space.

\subsection{Linear Koopman Modeling }
Direct computation with the infinite-dimensional Koopman operator $\mathcal{K}$ is generally intractable. However, when a finite set of \( \mathbf{q} \) observable functions is specified, the infinite-dimensional Koopman can be approximated by a finite-dimensional linear operator $\textbf{K}$ acting within the subspace spanned by the observable set. The original state-input pair \((\mathbf{x}, \mathbf{u})\) is thus lifted into a higher-dimensional embedding space through a vector-valued observable $\Psi(\mathbf{x}_k,\mathbf{u}_k)=[\psi^1(\mathbf{x}_k,\mathbf{u}_k), \psi^2(\mathbf{x}_k,\mathbf{u}_k), \dots,\psi^\mathbf{q}(\mathbf{x}_k,\mathbf{u}_k) ]^\top$ with $\psi^i:\mathbb{R}^n\times \mathbb{R}^m \to \mathbb{R}$, which satisfies 

\begin{equation}
    \Psi(\mathbf{x}_{k+1},\mathbf{u}_{k+1}) = \textbf{K}\Psi(\mathbf{x}_{k},\mathbf{u}_{k}).
    \label{eq:koopmanmatrix}
\end{equation}
To facilitate the control design, the vector-valued observable function $\Psi(\mathbf{x}_k, \mathbf{u}_k)$ can be further divided into two components
% However, to avoid causality problems regarding lifting the control inputs to a higher dimension subspace \cite{selby2021physics}, the observable function \( \Psi(\mathbf{x}_k,\mathbf{u}_k) \) can be divided into two components, which preserves linear dependency on inputs in the Koopman structure. Then the observable function can be characterized as
\begin{equation}
    \Psi(\mathbf{x}_k, \mathbf{u}_k) = \begin{bmatrix}
    \Psi_\mathbf{x}(\mathbf{x}_k) \\ 
    \Psi_\mathbf{u}(\mathbf{x}_k,\mathbf{u}_k)
    \end{bmatrix},
    \label{eq:linear observables}
\end{equation} where 
\begin{align}
    \Psi_\mathbf{x}(\mathbf{x}_k) &=[\psi^1(\mathbf{x}_k), \ldots, \psi^\mathbf{p}(\mathbf{x}_k)]^\top \in \mathbb{R}^\mathbf{p}, \\
    \Psi_\mathbf{u}(\mathbf{x}_k,\mathbf{u}_k) &= [\psi^{\mathbf{p}+1}(\mathbf{x}_k,\mathbf{u}_k), \ldots, \psi^{\mathbf{q}}(\mathbf{x}_k,\mathbf{u}_k)]^\top \in \mathbb{R}^\mathbf{d},
\end{align}
with \( \mathbf{d} = \mathbf{q} - \mathbf{p} \), where $\mathbf{p}$ denotes the number of state-dependent observables. For a causal linear Koopman realization, it is assumed $\Psi_\mathbf{u}(\mathbf{x}_k,\mathbf{u}_k) = \mathbf{u}_k$, i.e., the state-input dependent observable component is dependent only on inputs \cite{vicente2020linear}.
Then the system dynamics under the approximated Koopman matrix $\textbf{K}$ can be described as
\begin{equation}
\begin{bmatrix}
\Psi_\mathbf{x}(\mathbf{x}_{k+1}) \\
\mathbf{u}_{k+1}
\end{bmatrix}
=
\begin{bmatrix}
A_{xx} & B_{xu} \\
A_{ux} & B_{uu}
\end{bmatrix}
\begin{bmatrix}
\Psi_\mathbf{x}(\mathbf{x}_k) \\
\mathbf{u}_k
\end{bmatrix}.
\label{eq:lifted_dynamics_2}
\end{equation}
Since the focus is solely on the evolution of observables associated with the system states, and the control inputs are treated as externally specified by the controller rather than dynamically modeled, the representation reduces to a form governed solely by the state-dependent observable dynamics, as
\begin{equation}
\Psi_\mathbf{x}(\mathbf{x}_{k+1}) = A_{xx} \Psi_\mathbf{x}(\mathbf{x}_k) + B_{xu} \mathbf{u}_k,
\label{eq:koopman_linear}
\end{equation}
where $A_{xx} \in \mathbb{R}^{\mathbf{p} \times \mathbf{p}}$ and $ B_{xu} \in \mathbb{R}^{\mathbf{p} \times m} $ are the finite-dimensional Koopman approximation.
\subsection{Bilinear Koopman Modeling } \label{Bilinear Koopman Modeling}

In linear Koopman modeling, control-related observables are typically chosen directly as the control inputs. However, a more general form can be expressed as \( \Psi_\mathbf{u}(\mathbf{x}_k, \mathbf{u}_k) = [\mathbf{u}_k^\top, h(\mathbf{x}_k, \mathbf{u}_k)^\top]^\top \), where \( h(\cdot, \cdot) \) denotes a vector of nonlinear functions of states and inputs. While this generalization may improve model fidelity, it introduces nonlinearities with respect to control inputs and may result in non-causal system dynamics if the values of these observables cannot be directly measured \cite{selby2021physics}. Such nonlinear dependencies compromise one of the main advantages of Koopman-based approaches, which is preserving linearity to facilitate efficient and computationally tractable control design.
To strike a balance between model accuracy and optimization tractability, recent studies have proposed bilinear Koopman realizations \cite{bruder2021advantages, wang2024deep, zhao2024deep}. Bilinear terms maintain compatibility with Koopman frameworks while offering enhanced expressiveness compared to purely linear models.
The bilinear interaction term \(h(\mathbf{x}_k,\mathbf{u}_k) \) can be specified as a Kronecker product of the control input vector and state-dependent observable vector such that 
    \begin{align}
    % % \Psi_\mathbf{u}(\mathbf{x},\mathbf{u}) &= \begin{bmatrix}
    % %     \mathbf{u}\\
    % %     h(\mathbf{x},\mathbf{u})
    % \end{bmatrix}\\
        h(\mathbf{x}_k,\mathbf{u}_k)&= \mathbf{u}_k \otimes \Psi_\mathbf{x}(\mathbf{x}_k) = 
        \begin{bmatrix}
            u^1_k \cdot \Psi_\mathbf{x}(\mathbf{x}_k)  \\
            \vdots\\
             u^m_k \cdot \Psi_\mathbf{x}(\mathbf{x}_k) 
        \end{bmatrix}, 
    \end{align}
where \( u^{i}_k \) denotes the \( i \)-th component of the control input vector \( \mathbf{u}_k\), for \( i = 1, \dots, m \). This structure can then be utilized to augment the vector-valued observable defined in (\ref{eq:linear observables}), yielding an extended observable vector given by
\begin{equation}
    \Psi(\mathbf{x}_k, \mathbf{u}_k)=
    \begin{bmatrix}
        \Psi_\mathbf{x}(\mathbf{x}_k)\\
        \mathbf{u}_k\\
        u^1_k \cdot \Psi_\mathbf{x}(\mathbf{x}_k)\\
        \vdots\\
        u^m_k \cdot \Psi_\mathbf{x}(\mathbf{x}_k)
    \end{bmatrix}.
\end{equation} 
% Although this simplification facilitates efficient computation and preserves convexity in downstream optimization tasks such as model predictive control, it may limit the model's accuracy particularly when the system exhibits strong nonlinear interactions between states and inputs.

Finally, the bilinear Koopman representation can be expressed as
\begin{equation}
        \Psi(\mathbf{x}_{k+1},\mathbf{u}_{k+1}) =
    \begin{bmatrix}
        A_{xx} & B_{xu} & H_1 & \cdots &H_m \\
        \vdots & \vdots & \vdots  & \ddots& \vdots
    \end{bmatrix}
    \Psi(\mathbf{x}_k, \mathbf{u}_k),
    \label{eq:lifted_dynamics}
\end{equation}
where the evolution of state-dependent observables is explicitly described by
\begin{equation}
    \Psi_\mathbf{x}(\mathbf{x}_{k+1}) = A_{xx} \Psi_\mathbf{x}(\mathbf{x}_k) + B_{xu} \mathbf{u}_k + \sum_{i=1}^m H_i \left[u^i_{k} \cdot  \Psi_\mathbf{x}(\mathbf{x}_k) \right].
    \label{eq:bilinear_representation}
\end{equation}
 Here, the matrices  \( A_{xx} \) and \( B_{xu} \) represent the linear dynamics in the lifted space, and remain constant over time, while the time invariant bilinear matrices \( H_i \in \mathbb{R}^{\mathbf{p} \times \mathbf{p}} \) encode the multiplicative interactions between each control input channel $u^i_{k}$ and the state dependent observable features $\Psi_\mathbf{x}(\mathbf{x}_k)$. Such a bilinear formulation preserves linearity with respect to decision variables within an MPC framework, as will be discussed in Section~\ref{sec:CER-MDBK_MPC}, while enabling a more accurate modeling of control-dependent dynamics. Notably, when all \( H_i, \, i=1,\dots,m \), are zero, the model reduces to the standard linear Koopman realization.
 
To improve clarity, a slight abuse of notation was adopted by denoting \( A := A_{xx} \) and \( B := B_{xu} \) throughout the remainder of the paper. The matrices \( A \), \( B \), and \( H_i \)'s must be identified from data, alongside the state-dependent lifting functions. These lifting functions can either be selected from a predefined library of observables or learned directly using deep neural network (DNN) architectures. The lifted states are denoted by \( Z_k = \Psi_{\mathbf{x}}(\mathbf{x}_k) \), where \( \Psi_{\mathbf{x}}(\cdot) \) represents the vector of state-dependent observable functions that map the original state space to the lifted subspace. Building on this theoretical foundation, the following section presents a comprehensive methodology for data-driven identification of these lifting functions and Koopman matrices, which ultimately leads to their integration within a predictive control framework.

%%%%%%%%%%%%%%%%%%%%%%%%%%%%%%%%%%%%%%%%%%%%%%%%%%%%%%%%%%%%%%%%%%
\section {Methodology} \label{sec:Methodology}
%%%%%%%%%%%%%%%%%%%%%%%%%%%%%%%%%%%%%%%%%%%%%%%%%%%%%%%%%%%%%%%%%%
\begin{figure*}[t]
    \centering
    \includegraphics[width=\textwidth]{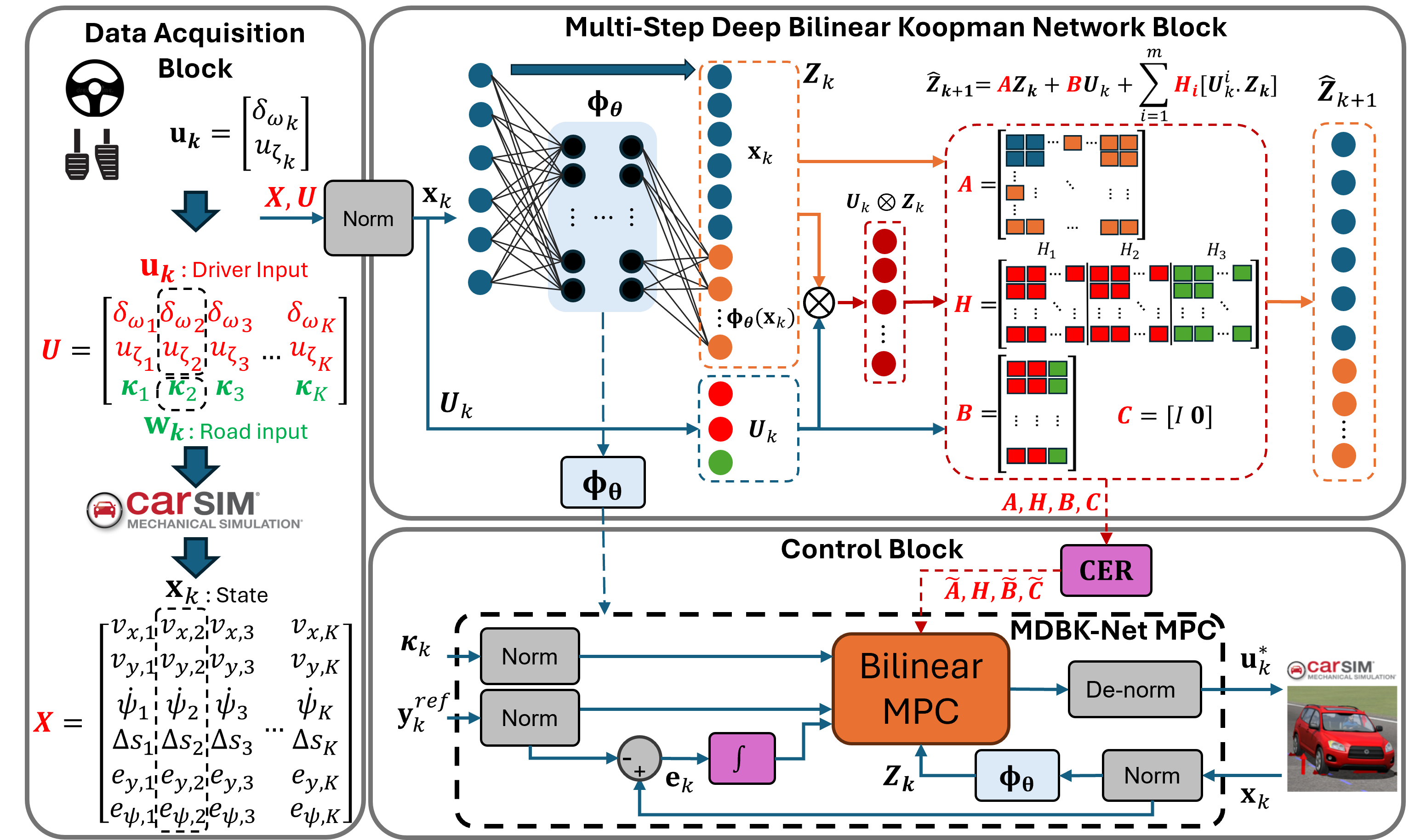}
    \caption{Architecture of the proposed MDBK-Net framework. The system consists of three main components: (i) a data acquisition block for preparing training and test datasets, (ii) a multistep Deep Bilinear Koopman Network block that defines the neural architecture used for learning the Koopman model, and (iii) a control block that employs the identified observables and Koopman dynamics as the predictive model within an optimal control framework. The control block is further enhanced with the CER to improve tracking performance. In this figure, Norm stands for data normalization, and De-norm stands for de-normalizing the data to its original values.}
    \label{fig:network_architecture}
\end{figure*}
In this section, the nonlinear vehicle dynamics that need to be identified using the bilinear Koopman framework are first presented. For identification and performance evaluation, a structured three-stage pipeline was adopted. In the first stage, a rich and diverse dataset is generated using the CarSim simulation platform under varying initial conditions and driving scenarios to accurately capture a broad range of the vehicle’s dynamic behavior. In the second stage, the MDBK-Net architecture was introduced, which jointly learns the lifting functions and the Koopman matrices. To mitigate long-horizon prediction drift caused by model approximation errors, the learned Koopman model is augmented with the CER block. In the final stage, the model's performance is evaluated through open-loop simulations, while real-time closed-loop performance is assessed under an MPC scheme using hardware-in-the-loop (HIL) testing.
An overview of the complete modeling and control pipeline, which includes data acquisition, encoder and Koopman matrix identification, integration with the CER module, and MPC deployment, is illustrated in Fig.~\ref{fig:network_architecture}.

\subsection{Vehicle Dynamics}

% \subsubsection{General Dynamics Formulation}
To establish a foundation for subsequent discussions and to gain insight into the key nonlinearities present in the vehicle dynamics illustrated in Fig.~\ref{fig:fernet frame}, a simplified nonlinear bicycle model is introduced, formulated in the Frenet frame as 
\begin{subequations} \label{eq:frenet_equations}
\begin{align}
\dot{v}_x &= \frac{1}{m} \left( F_x^f \cos(\delta^f) - F_y^f \sin(\delta^f) + F_x^r \right) + \dot{\psi} v_y \label{eq:frenet_vx} \\
\dot{v}_y &= \frac{1}{m} \left( F_x^f \sin(\delta^f) + F_y^f \cos(\delta^f) + F_y^r \right) - \dot{\psi} v_x \label{eq:frenet_vy} \\
\ddot{\psi} &= \frac{1}{I_z} \left( l_f F_x^f \sin(\delta^f) + l_f F_y^f \cos(\delta^f) - l_r F_y^r \right) \label{eq:frenet_psidd}
\end{align}
\begin{align}
\dot{s} &= \frac{v_x \cos(e_\psi) - v_y \sin(e_\psi)}{1 - \kappa e_y} \label{eq:frenet_s} \\
\dot{e}_y &= v_x \sin(e_\psi) + v_y \cos(e_\psi) \label{eq:frenet_ey} \\
\dot{e}_\psi &= \dot{\psi} - \frac{v_x \cos(e_\psi) - v_y \sin(e_\psi)}{1 - \kappa e_y} \kappa \label{eq:frenet_eps}
\end{align}
\end{subequations}
Here, \( v_x \), \( v_y \), \( \dot{\psi} \), \( s \), \( e_y \), and \( e_\psi \) denote the vehicle states as longitudinal velocity, lateral velocity, yaw rate, progress along the path, lateral deviation from the reference path, and heading error, respectively. The parameters \( m \) and \( I_z \) represent the vehicle's mass and yaw moment of inertia. The terms \( F_x^f, F_y^f, F_x^r, F_y^r \) correspond to the longitudinal and lateral tire forces at the front and rear wheels, \( \delta^f \) denotes the front wheel steering angle, while the variable \( \kappa \) represents the curvature of the reference path. Note that although the above equations of motion capture key nonlinearities, the actual vehicle dynamics are significantly more complex. The high-fidelity CarSim model used here includes steering system dynamics, powertrain and braking systems, load transfer effects, and nonlinear tire behaviors. Accurate modeling of such a system within a control-oriented and optimization-friendly framework is often impractical in many real-world scenarios. This motivates the use of data-driven modeling techniques to identify precise representations of vehicle dynamics from observed behavior.

Accordingly, a general discrete-time nonlinear mapping of vehicle dynamics is considered as
\begin{equation}
\mathbf{x}_{k+1} = f_v(\mathbf{x}_{k}, \mathbf{u}_{k}, \mathbf{w}_{k}),
\label{eq:general_vehicle_dynamics}
\end{equation}
where $k$ is the discrete time index and $\mathbf{x}_k \in \mathbb{R}^n$,  $\mathbf{u}_k \in \mathbb{R}^m$, and $\mathbf{w}_k \in \mathbb{R}^l$ represent vehicle states, driver input commands, and road exogenous inputs, respectively. The function $f_v: \mathbb{R}^{n+m+l}\to \mathbb{R}^n$ denotes the nonlinear state transition mapping of the vehicle dynamics. Similar to Eq.~\eqref{eq:frenet_equations}, the state vector can be explicitly denoted by

\begin{equation}
\mathbf{x} = [v_x, v_y, \dot{\psi}, \Delta s, e_y, e_\psi]^\top,
\label{eq:state vector}
\end{equation}
with a slight modification, the term \( \Delta s \) represents the progress gain at each time step along the reference path. This modification is introduced to improve the consistency and numerical stability of the training process in the deep learning framework presented later.

Note that the representation in (\ref{eq:frenet_equations}) accepts the front-wheel steering angle and tire forces as system inputs; however, in the dynamical model (\ref{eq:general_vehicle_dynamics}), the control input vector corresponds to driver commands given by
\begin{equation}
\mathbf{u} = [\delta_w, u_\zeta]^\top,
\label{eq:input vector}
\end{equation}
where \( \delta_w \) denotes the hand wheel steering angle, and \( u_\zeta \in [-1,1] \) represents the longitudinal drive command, with positive values corresponding to throttle and negative values to braking. Additionally, the exogenous input vector, which captures the effect of road curvature on the vehicle’s pose, is defined as
\begin{equation}
\mathbf{w} = [\kappa]^\top.
\label{eq:exogenous_input_vector}
\end{equation}
To accurately identify and model the nonlinear dynamics described above within the Koopman operator framework, it is crucial to generate a comprehensive dataset that reflects the full range of vehicle behavior under realistic driving conditions. A high-fidelity dataset was generated using the CarSim simulation platform. The data collection process is described in the following section.

\subsection{Data Acquisition and Preparation}\label{sec:Data Acquisition}
The training dataset was generated using CarSim, a high-fidelity simulation platform for modeling real-world vehicle behavior. A C-Class hatchback vehicle configuration equipped with a 150 kW engine and an all-wheel-drive system was used. A total of 7,000 simulated episodes are generated across various driving scenarios, each lasting 10 seconds with a sampling interval of \( t_s = 25~\mathrm{ms} \), yielding 400 discrete time steps per episode. To further diversify the data, introduce varied initial conditions, and maintain consistency with the prediction horizon used in MPC, each episode is divided into five non-overlapping 2-second segments, resulting in 35,000 distinct trajectories.

Each simulated driving scenario captures the evolution of vehicle states governed by the applied control inputs. Rather than using arbitrary inputs, a structured input generation procedure is employed to ensure realistic driving behavior. Exogenous road curvature values \( \kappa \) are sampled uniformly from the range \( [-4 \times 10^{-3}, 4 \times 10^{-3}] \,\text{m}^{-1} \),  corresponding to turning radius as small as 250 meters in either direction.

At each time step \( k = 1, \ldots, 400 \), the driver command \( u_k \) is constructed such that the steering wheel input \( \delta_w \in [-40^\circ, 40^\circ] \) is generated by fitting a polynomial curve to uniformly distributed key points across the episode horizon, producing a smooth steering profile. The longitudinal actuation input \( u_\zeta \in [-1, 1] \) is simulated, where positive values denote throttle and negative values represent braking effort, scaled proportionally to reflect pedal force. This normalization enforces mutual exclusivity between throttle and brake commands, ensuring that the brake input is zero whenever the throttle is active, and vice versa. Longitudinal drive command rates are constrained to reflect realistic driver behavior, guided by human reaction time estimates from \cite{matheson2019study}. This is achieved by segmenting each episode into one-second intervals, which promotes consistent pedal usage within each segment.

The synthesized input sequence is applied to the CarSim vehicle model, and the resulting state trajectories are recorded. All inputs and states are subsequently normalized using the mean and standard deviation computed across the entire dataset, denoted by Norm blocks as shown in Fig. \ref{fig:network_architecture}. Finally, the resulting dataset serves as the basis for training the MDBK Network.

\subsection{MDBK Network Architecture}
The MDBK-Net introduced in this work explicitly modeled both linear and bilinear interactions between lifted states and control inputs. In the following subsections, each component of the MDBK-Net architecture is detailed.

\subsubsection{Encoder Network Structure}
Unlike traditional approaches where the lifting functions are selected from predefined libraries of nonlinear basis functions, such as polynomials, or radial basis functions, a data-driven strategy is employed using neural networks \textit{to learn the lifting functions directly from data}. 

As illustrated in Fig.~\ref{fig:network_architecture}, an encoder network \( \Phi_\theta \), parameterized by weights \( \theta \), maps the original state \( \mathbf{x}_k \in \mathbb{R}^n \) to a subspace \( \Phi_\theta(\mathbf{x}_k) \in \mathbb{R}^{p - n} \). The lifted components are given by \(  \Phi_\theta(\mathbf{x}_k) = [\phi^1(\mathbf{x}_k), \ldots, \phi^{p-n}(\mathbf{x}_k)]^\top \), where each \( \phi^i : \mathbb{R}^n \rightarrow \mathbb{R} \) denotes a nonlinear function learned by the neural network. The encoder network is implemented as a fully connected feedforward multi-layer perceptron (MLP) utilizing ReLU activation functions, with specific configurations detailed in Table~\ref{tab:hyperparams}. The dimension $p$  of the lifted subspace is treated as a design parameter; increasing \( p \) generally enhances model fidelity but also increases computational cost and the risk of overfitting. Evaluation of different network architectures indicated that using 60 nodes in the encoder’s output layer offers the best trade-off between accuracy and model complexity.
To facilitate interpretability and accurate reconstruction of the original states, the lifted space \( Z_k \) is constructed by concatenating the original state \( \mathbf{x}_k \) with the encoder output layer. Accordingly, the full lifted state space is given by
\begin{equation}
    Z_k = \begin{bmatrix}
        \mathbf{x}_k \\ \Phi_\theta(\mathbf{x}_k) 
    \end{bmatrix} \in \mathbb{R}^{p}.
    \label{eq: vector of observables}
\end{equation}
This structure guarantees that the predicted original states can be later recovered using a simple linear projection matrix.

Given the discrete-time vehicle dynamics in (\ref{eq:general_vehicle_dynamics}), The combined input vector is defined, which applies to the vehicle dynamics as
\begin{equation}
    U_k =
    \begin{bmatrix}
        \mathbf{u}_k\\
        \mathbf{w}_k
    \end{bmatrix}\in \mathbb{R}^{m + l},
    \label{eq:control_input_vetor}
\end{equation}
where \( U_k \) is the concatenation of the driver control inputs \( \mathbf{u}_k \) and road exogenous inputs \( \mathbf{w}_k \). Note that road specifications such as curvature \( \kappa \) are independent of the vehicle state and are typically available through onboard sensors or perception systems. For further details on incorporating such exogenous inputs, the reader is referred to \cite{kim2023koopman} \cite{kim2025k}.
The lifted states \( Z_k \), the bilinear interaction term \( ( U_k \otimes Z_k )^\top \), and the control input \( U_k \) are then passed to the final Koopman layer of the network.
\subsubsection{Koopman Layer Structure}
The Koopman layer is implemented as the last neural network layer without any bias or activation functions, where the Koopman operator is approximated using learnable matrices. The matrices \( A \), \( B\), and  the set of bilinear matrices \( \{H_i\}_{i=1}^{m+l} \) will be constructed and trained simultaneously with encoder parameters.
Then the corresponding single-step bilinear Koopman propagation based on (\ref{eq:bilinear_representation}) can be expressed as 
\begin{align}
   \hat{Z}_{k+1} &= AZ_{k} + B U_k + \sum_{i=1}^{m+l} H_i \left[  U^i_{k}\cdot Z_{k}  \right], \label{eq:deep_bilinear_koopman_modeling} 
\end{align}
where $Z_{k}$ denotes the lifted state vector constructed according to (\ref{eq: vector of observables}) and $\hat{Z}_{k+1}$ represents the predicted lifted state at the next time step. The matrix \( A \in \mathbb{R}^{p \times p} \) represents the Koopman vehicle dynamics, and the input matrix \( B=[B_\mathbf{u}, B_\mathbf{w}] \in \mathbb{R}^{p \times (m + l)} \) maps the combined vehicle inputs to the next step. This matrix is composed of \( B_\mathbf{u} \in \mathbb{R}^{p \times m} \) corresponds to driver commands channel and \( B_\mathbf{w} \in \mathbb{R}^{p \times l} \) corresponds to exogenous road inputs (curvature) channel.
Each matrix \( H_i \in \mathbb{R}^{p \times p} \) corresponds to the \( i \)-th input channel \( U^i_k \) of the combined input vector, making the bilinear term \( \sum_{i=1}^{m+l} H_i \left[ U^i_k\cdot Z_k  \right] \) an input-specific correction to the lifted dynamics.

\subsubsection{Original States Reconstruction}
Based on the lifting structure introduced in (\ref{eq: vector of observables}), the projection operation used to recover the original system states from the predicted lifted states is given by
\begin{equation}
    \label{eq:projection}
    \hat{\mathbf{x}}_k = C \hat{Z}_k,
\end{equation}
where the projection matrix \( C \in \mathbb{R}^{n \times p} \) is given by
\begin{equation}
    C = [\textbf{\textit{I}}_n \quad \textbf{0}],
    \label{eq:projection_matrix}
\end{equation}
with \( \textbf{\textit{I}}_n \) denoting the \( n \times n \) identity matrix and \( \textbf{0} \in \mathbb{R}^{n \times (p - n)} \) denoting a zero matrix. This projection guarantees that the
original vehicle states are exactly embedded in the lifted space and can be retrieved without approximation.

\subsection{Training Procedure and Loss Functions}
The lifting functions and Koopman matrices are jointly learned with the encoder parameters, Koopman transition matrices, and bilinear interaction terms in an end-to-end deep learning framework using backpropagation.
Training is performed using batched gradient descent over multiple epochs. Each batch consists of a fixed number of trajectories randomly sampled from the normalized dataset. During each iteration, the encoder maps the states to a lifted state space \( \Phi_\theta(\mathbf{x}_k) \). The lifted states vector \( Z_k \) is then constructed by concatenating \( \Phi_\theta(\mathbf{x}_k) \) with the original state vector \( \mathbf{x}_k \), and is subsequently used to predict the next lifted state \( \hat{Z}_{k+1} \) via the Koopman transition.

An essential aspect of the training process is designing a well-structured loss function that effectively guides the learning of the model dynamics. Unlike traditional Koopman identification techniques that focus exclusively on single-step prediction, the proposed MDBK-Net adopts a comprehensive loss structure that balances short-term and long-term prediction accuracy, following the loss formulation proposed in  \cite{xiao2023ddk} and \cite{lusch2018deep}. Additionally, to ensure stability of the learned dynamics and to prevent overfitting, the framework incorporates both a spectral stability loss and a regularization term.
For each training trajectory of length \( K \), the loss components are defined as follows.

The \textbf{single-step loss} \( \mathcal{L}_{\text{ssl}} \) penalizes the deviation between the predicted  \( \hat{Z}_{k+1} \), computed using (\ref{eq:deep_bilinear_koopman_modeling}), and the ground truth \( Z_{k+1} \) constructed based on (\ref{eq: vector of observables}), at each time step
    \begin{equation}
    \begin{aligned}
    \mathcal{L}_{\text{ssl}} = \frac{1}{K-1} \sum_{k=1}^{K-1} \Big\| Z_{k+1} - \hat{Z}_{k+1}
    % - \big( A \Psi_\mathbf{x}(\mathbf{x}_k) 
    % + B \mathbf{U}_k \\
    % + \sum_{i=1}^{m} H_i [ \Psi_\mathbf{x}(\mathbf{x}_k) \cdot \mathbf{U}_{k,i} ] \big) 
    \Big\|_2^2.
    \end{aligned}
    \end{equation}
To improve long-horizon predictive capabilities, a \textbf{multi-step prediction loss} \( \mathcal{L}_{\text{msl}} \) is introduced. This loss recursively predicts the lifted states and penalizes accumulated error across the horizon using a forgetting factor \( \beta \in (0,1) \) 
    \begin{equation}
    \begin{split}
        \mathcal{L}_{\text{msl}} &= \frac{1}{\sum_{k=1}^{K-1} \beta^k} \sum_{k=1}^{K-1} \beta^k \left\| Z_{k+1} - \hat{Z}^{rec}_{k+1} \right\|_2^2,
    \end{split}
    \label{eq:bilinear_msl}
    \end{equation}
where the recursive predictions $\hat{Z}_{k+1}^{rec}$ is computed by
    \begin{equation}
    \begin{split}
        \hat{Z}^{rec}_{k+1} &= A \hat{Z}^{rec}_{k} + B U_{k} 
        + \sum_{i=1}^{m+l} H_i \left[ U^i_{k} \cdot\hat{Z}^{rec}_{k}  \right], \\
        \hat{Z}^{rec}_{1} &= Z_1,
    \end{split}
    \end{equation}
and $Z_{k}$ constructed by (\ref{eq: vector of observables}).

To enforce stability in the learned dynamics, a \textbf{stability loss} \( \mathcal{L}_{\text{sl}} \) is applied to penalize eigenvalues of the Koopman matrix \( A \)  that lie outside the unit circle.
    \begin{equation}
        \mathcal{L}_{\text{sl}} = \sum_{\lambda \in \text{eig}(A)} \max(0, |\lambda| - 1).
        \label{eq:bilinear_sl}
    \end{equation}
To promote generalization and prevent overfitting, a \textbf{regularization loss} $\mathcal{L}_{\text{reg}}$ is imposed on the encoder parameters and Koopman matrices. 
% Regularization is imposed on the encoder weights, as well as on the Koopman matrices \( A, B \), and the bilinear terms \( \{H_i\}_{i=1}^{m+l} \), in order to promote generalization and avoid overfitting. The regularization loss  \( \mathcal{L}_{\text{reg}} \) is formulated as
    \begin{equation}
    \mathcal{L}_{\text{reg}} = 
    \lambda_{\theta} \sum_{\theta_j \in \theta} \|\theta_j\|^2_2 
    + \lambda_{\text{AB}} \left( \|A\|_F^2 + \|B\|_F^2 \right)
    + \lambda_{H} \sum_{i=1}^{m+l} \|H_i\|_F^2,
    \label{eq:reg_loss}
    \end{equation}
where \( \|\theta_j\|_2 \) denotes the \( \ell_2 \) regularization applied to the encoder parameters, and \( \|\cdot\|_F \) represents the Frobenius norm used to regularize the Koopman matrices. The terms \( \lambda_{\theta} \), \( \lambda_{\text{AB}} \), and \( \lambda_H \) are regularization weights corresponding to the encoder parameters, linear Koopman matrices, and bilinear interaction terms, respectively.

The total \textbf{training loss} is defined as a weighted combination of the above components
\begin{equation}
    \mathcal{L}_{\text{total}} = \alpha_1\,\mathcal{L}_{\text{ssl}} 
    + \alpha_2\,\mathcal{L}_{\text{msl}} 
    + \alpha_3\,\mathcal{L}_{\text{sl}} 
    + \alpha_4\,\mathcal{L}_{\text{reg}},
    \label{eq:total_loss}
\end{equation}
where the weights \( \alpha_1, \alpha_2, \alpha_3, \alpha_4 \) control the contribution of each loss component during training. To identify the optimal network architecture and set of optimization hyperparameters, multiple configurations of the model were trained and evaluated. The best-performing configuration was selected based on validation performance. All associated hyperparameters, including regularization coefficients and loss weights, are summarized in Table~\ref{tab:hyperparams}.

\begin{table}[b]
\centering
\caption{MDBK-Net Architecture and  Training Hyperparameters}
\label{tab:hyperparams}
\begin{tabular}{ll}
\hline
\textbf{Hyperparameter} & \textbf{Value} \\
\hline
Max Number of Epochs ($Epoch_{\max}$) & 80,000 \\
Data Split (Train/Test) & 90\%/10\%\\
Batch Size & 128 \\
Observables State Space Dimension (\(p\)) & 66 \\
Hidden Layer Sizes & [32, 64, 128, 128, 64] \\
Initial Learning Rate ($lr_\text{init}$) & $10^{-3}$ \\
Min Learning Rate ($lr_\text{min}$) & $5\times10^{-7}$ \\
Learning Rate Reduction factor ($\mu$) & 0.5\\
Patience& 2\\
Forgetting  Factor (\(\beta\)) & 0.9 \\
Loss Function Weights (\(\alpha_1,\ldots,\alpha_4\)) & [0.1, 1.0, 1.6, $10^{-4}$] \\
Regularization Factors (\(\lambda_\theta,\lambda_{AB},\lambda_{H}\)) & 10, 1, 100\\
State Dimension (\(n\)) & 6 \\
Driver Input Dimension (\(m\)) & 2 \\
Exogenous Input Dimension (\(l\)) & 1 \\
Time Step (\(\Delta t\)) & 0.025\,s \\
\hline
\end{tabular}
\end{table}
\begin{algorithm}
\caption{MDBK-Net Algorithm}
\label{alg:MDK-Net}
\begin{algorithmic}[1]
\State Initialize Model Parameteres $\theta$, $A$, $B$, $H$,
\State Initialize $Epoch_{\max}$, batch size, data split, patience, $lr$, $\beta$, $\{\alpha_i\}_{i=1}^4$,  Validation total loss vector $\mathcal{L}^{\text{val}}$, best model $\Gamma$, $i\gets 0$.
\State Randomly partition the data into train and test sets and normalize them.
\While {$Epoch < Epoch_{\max}$}
    \State Randomly select a batch.
    \State Encode the state vector and concatenate it with original states as in (\ref{eq: vector of observables}).
    \State Compute total loss as in (\ref{eq:total_loss}) and its gradient.
    \State Update $\theta$, $A$, $B$, $H$ using the computed gradient.
    \If {$Epoch$ $\mathbf{mod}\,500 == 0$}
        \State Increment $i$ by 1.
        \State Evaluate total loss (\ref{eq:total_loss}) on the entire validation set and append it to $\mathcal{L}^{\text{val}}$.
        \If {$\mathcal{L}^{\text{val}}_i > \mathcal{L}^{\text{val}}_{i-1}>\dots>\mathcal{L}^{\text{val}}_{i-{\text{patience}}}$}
            \State Update the learning rate: $lr \gets lr \times \mu$.
            \If  {$lr = lr_\text{min}$}
                \State Break;
            \EndIf
        \ElsIf {$\mathcal{L}^{\text{val}}_{i} < \mathcal{L}^{\text{val}}_{i-1}$}
            \State Update best model: $\Gamma \gets\{\theta, A, B, H\}$.
        \EndIf
    \EndIf
    \State Increment $Epoch$ by 1.
\EndWhile
\State \textbf{Output:} Best model parameters $\Gamma=\{\theta, A, B, H\}$ 
\end{algorithmic}
\end{algorithm}
After computing the total loss at each iteration, backpropagation is performed to update both the encoder parameters \( \theta \) and the Koopman matrices \( A \), \( B \), and \( \{H_i\}_{i=1}^{m+l} \).
The training loop includes periodic evaluations of loss on a validation set to monitor model performance. If no improvement is observed within a predefined patience window, the initial learning rate \( lr_{\text{init}} \) is reduced by a factor of \( \mu \) to facilitate convergence, continuing until it reaches the minimum threshold \( lr_{\text{min}} \). The model achieving the lowest validation loss is retained as the final configuration based on an early stopping criterion, and the maximum number of epochs $Epoch_{\max}$ is set accordingly. The MDBK-Net training process, including learning schedule, loss evaluation, and optimization criteria, is summarized in Algorithm~\ref{alg:MDK-Net}, with hyperparameters listed in Table~\ref{tab:hyperparams}.

\subsection{MDBK-Net Performance for State Prediction Evaluation}
The performance of the MDBK-Net was evaluated against two baseline models:
(i) the MDK-Net algorithm as a linear Koopman-based counterpart, and
(ii) a model identified using the classical Extended Dynamic Mode Decomposition with Koopman (EDMDK) technique, as described in \cite{cibulka2019data}, \cite{kim2023koopman}, and \cite{vsvec2024koopman}.
Generally, in the EDMDK framework, lifting functions are either manually specified based on prior knowledge of the system dynamics or empirically chosen by testing various basis functions. Following the approach in \cite{kim2023koopman}, a polynomial basis consisting of all monomials up to degree two was used to construct the lift functions for the EDMDK baseline. To ensure a fair comparison, the dimensionality of the lifted space in EDMDK was aligned with that of MDK-Net and MDBK-Net. All models were trained on the same dataset, and identical hyperparameter settings were used for both MDK-Net and MDBK-Net during training.

\begin{figure}[t]
    \centering
    \includegraphics[width=\linewidth]{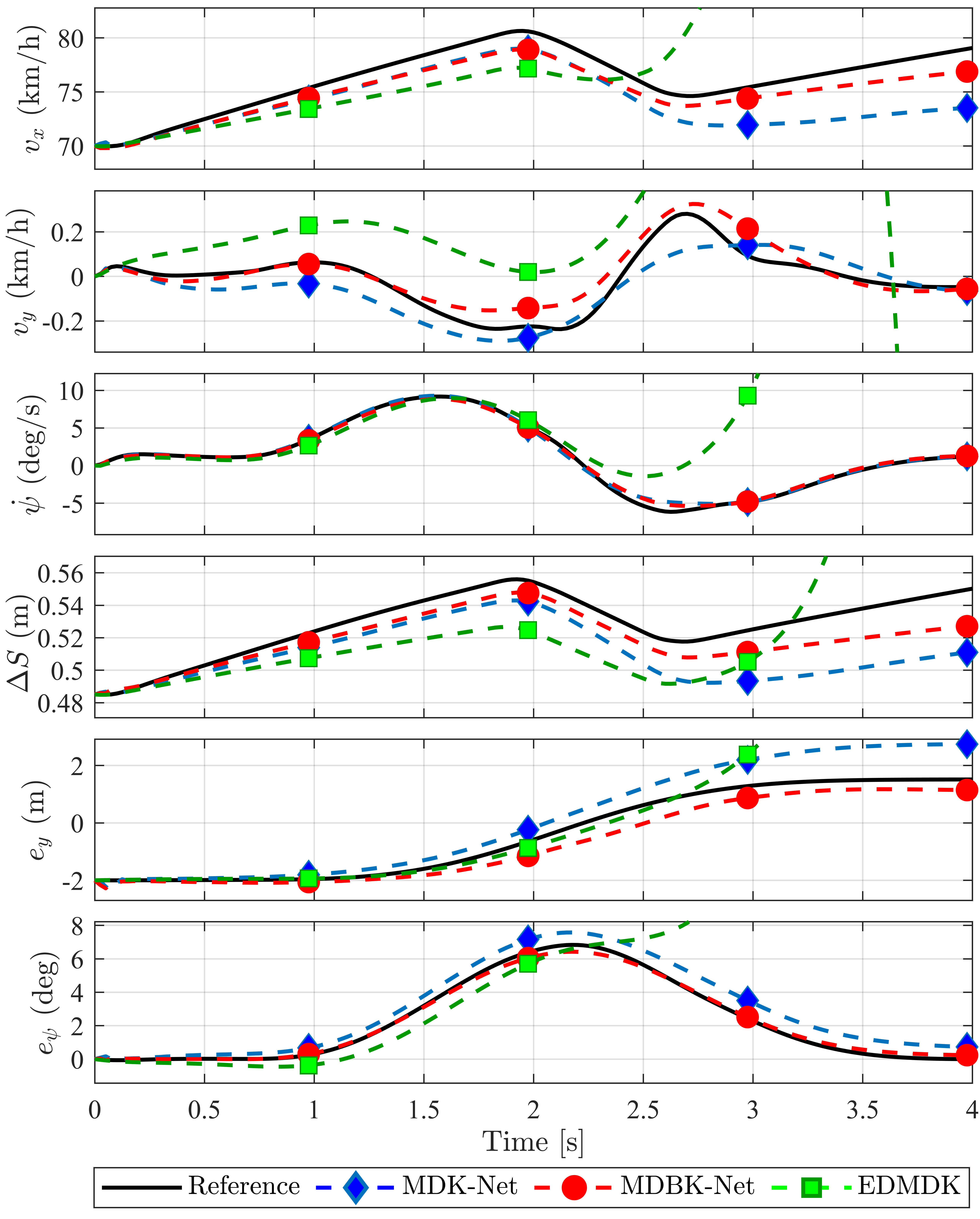}
    \caption{Long-term prediction results of vehicle dynamics for proposed MDBK-Net, MDK-Net, and EDMDK. The reference trajectory for all states are extracted from high-fidelity Carsim simulations, and the same control input sequence is applied to all identified models. The EDMDK predictions (green curves) exhibit divergence due to instability in the identified Koopman matrices, resulting in values that extend beyond the vertical axis limits.}
    \label{fig:models comparison}
\end{figure}

Figure \ref{fig:models comparison} illustrates the state prediction results for all three models under the same open-loop control input sequence. The ground-truth trajectories were obtained from CarSim simulations for a high-speed single-lane change scenario, during which the driver brakes mid-maneuver. Although the EDMDK algorithm was able to capture the dynamics of yaw rate (\(\dot{\psi}\)) and lateral deviation (\(e_y\)) up to a 2-second prediction horizon, it exhibits rapidly growing errors thereafter, indicating instability in the identified model. It also performs poorly in predicting longitudinal and lateral velocities (\(v_x\), \(v_y\)), progress gain (\(\Delta s\)), and heading error (\(e_\psi\)). The modeling inaccuracy is largely attributed to the limited expressiveness of manually selected observable functions, which lack the flexibility needed to capture a rich Koopman-invariant subspace. As a result, the identified Koopman model has become unstable, even with the implementation of Tikhonov regularization during training. This regularization technique introduces an \( \ell_2 \) penalty on the matrix coefficients to discourage large values and improve numerical stability \cite{golub1999tikhonov}.

In contrast, both MDK-Net and MDBK-Net learn lifting functions through deep neural encoders, which results in superior performance over EDMDK across nearly all state predictions. Their multi-step prediction capability further enhances their ability to capture temporal dependencies, leading to more accurate long-horizon predictions. In particular, the MDBK-Net model achieves superior accuracy in capturing system dynamics compared to the baseline models. This performance gain is primarily attributed to the model's ability to encode bilinear interactions between control inputs and lifted states, which will be discussed in detail through the analysis of the bilinear coefficient matrices. To quantify prediction accuracy, the root mean square error (RMSE) was evaluated across test trajectories, each spanning an 80-step (2-second) horizon with varying initial conditions. The 2-second window was selected to ensure fairness, as EDMDK often becomes unstable beyond this horizon, resulting in divergent predictions and invalid error metrics. It is important to note that all RMSE values are reported after normalizing each state using the corresponding mean and standard deviation computed over the entire test dataset.

As reported in Table~\ref{tab:rmse_comparison}, the MDBK-Net consistently achieves the lowest RMSE across all six state variables. The most significant improvements are observed in the lateral dynamics, where MDBK-Net reduces the prediction error in lateral velocity (\(v_y\)) by 43\% and yaw rate (\(\dot{\psi}\)) by 36\% compared to MDK-Net, and by 66\% and 50\%, respectively, relative to EDMDK. Furthermore, the proposed model demonstrates notable gains in curvilinear tracking states, with RMSE reductions of 22\% and 12\% in progress gain (\(\Delta s\)) and lateral deviation (\(e_y\)), and 40\% in heading error (\(e_\psi\)) relative to MDK-Net. While MDK-Net performs comparably to MDBK-Net in predicting longitudinal velocity (\(v_x\)), it underperforms in capturing lateral dynamics and curvilinear positioning. This performance gap highlights the importance of explicitly modeling bilinear interactions between driver inputs and road geometry, as done in MDBK-Net. By capturing these interactions, MDBK-Net provides a more accurate representation of complex lateral behavior, particularly under high-speed or aggressive driving conditions, thereby validating the benefits of bilinear Koopman modeling.

\begin{table}[t]
    \centering
    \setlength{\tabcolsep}{4pt}
    \caption{Average open-loop 80-step (2-second) prediction RMSE for 1,425 test trajectories}
    \label{tab:rmse_comparison}
    \begin{tabular}{lccc}
        \toprule
        \textbf{State Variable} & \textbf{MDBK-Net} & \textbf{MDK-Net} & \textbf{EDMDK} \\
        \midrule
        \(v_x\) (km/h)       & \textbf{0.138} & 0.155 & 0.214 \\
        \(v_y\) (km/h)       & \textbf{0.274} & 0.483 & 0.821 \\
        \(\dot{\psi}\) (deg/s) & \textbf{0.064} & 0.100 & 0.114 \\
        \(\Delta s\) (m)     & \textbf{0.133} & 0.171 & 0.207 \\
        \(e_y\) (m)          & \textbf{0.071} & 0.081 & 0.121 \\
        \(e_\psi\) (deg)     & \textbf{0.043} & 0.072 & 0.114 \\
        \bottomrule
    \end{tabular}
\end{table}

To further understand how these bilinearities contribute to performance improvements, the next section analyzes the identified bilinear matrices and their role in modeling input-specific interactions within the Koopman framework.
\subsection{Sparsity Patterns of Bilinear Coefficient Matrices}

\begin{figure*}[t]
    \centering
    \includegraphics[width=\textwidth]{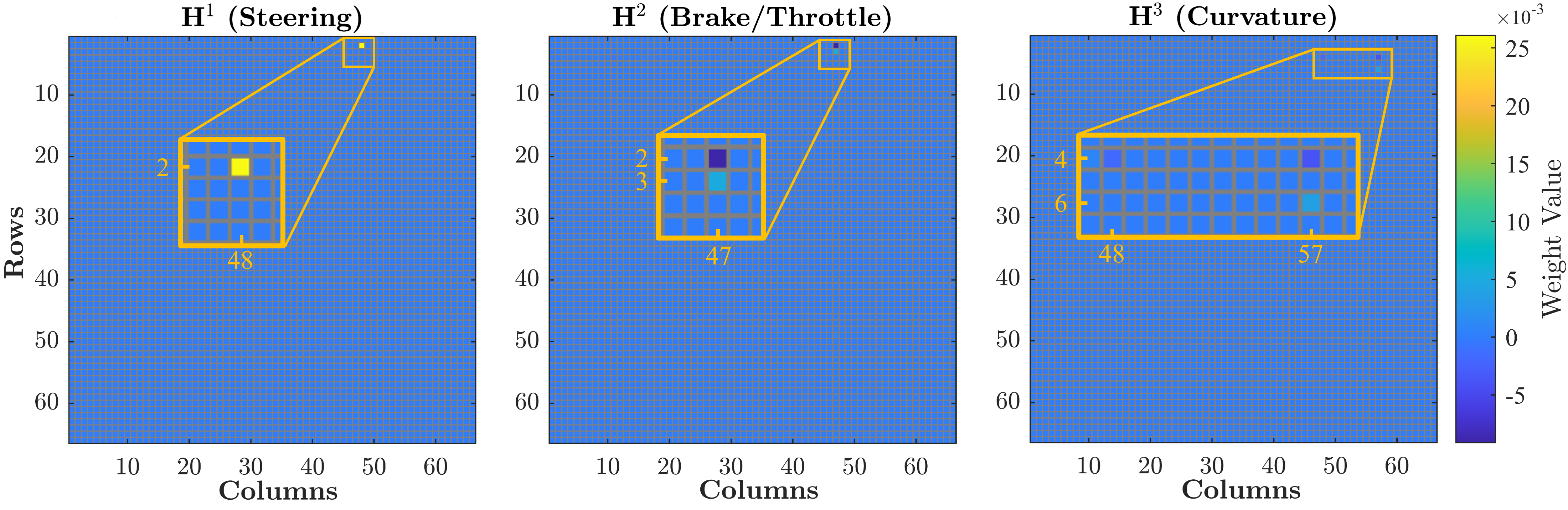}
    \caption{Visualization of bilinear Koopman matrices \( H_1 \), \( H_2 \), and \( H_3 \), corresponding to steering input, longitudinal drive command, and road curvature, respectively. Each matrix encodes input-specific multiplicative interactions with the observable vector. Sparsity patterns highlight how MDBK-Net captures key nonlinear dependencies across different control channels, particularly for lateral dynamics and curvilinear tracking states.}
    \label{fig:H matrices}
\end{figure*}

MDBK-Net's performance advantage can be attributed to its ability to encode bilinear interactions between all control inputs and lifted features. To investigate this, the sparsity patterns of the bilinear term coefficient matrices\(\{H_i\}_{i=1}^{3}\), which capture these interactions within the Koopman dynamics, were examined. As shown in Fig.~\ref{fig:H matrices}, the matrices \(H_1\) (associated with the driver steering input), \(H_2\) (longitudinal drive command), and \(H_3\) (road curvature) reveal noticeable bilinear dependencies influencing the evolution of states. 

The sparsity patterns observed in the bilinear matrices \(H_1\) and \(H_2\) suggest that the MDBK-Net algorithm effectively captures the influence of bilinear interactions on the lateral velocity \(v_y\). Specifically, a prominent entry in \(H_1\) linking the steering input to the 48th lifted state highlights a strong nonlinear coupling in the steering dynamics that governs lateral motion. This finding aligns with (\ref{eq:frenet_vy}), which shows that sinusoidal functions of the steering angle regulate lateral velocity. Similarly, \(H_2\) reveals interactions between the longitudinal drive input and the 47th lifted state, affecting both \(v_y\) and \(\dot{\psi}\), which correspond to the second and third lifted states. This observation aligns with the nonlinearities arising from tire slip dynamics under conditions approaching handling limits, as described by (\ref{eq:frenet_vy}) and (\ref{eq:frenet_psidd}), which affect both the lateral velocity and yaw rate. Finally, \(H_3\) uncovers bilinear contributions to the curvilinear tracking dynamics, where the exogenous road curvature input 
$\kappa$ interacts with the 48th and 57th lifted states. These interactions confirm that the progress gain $\Delta s$ and heading error $e_\psi$, corresponding to the fourth and sixth states in the lifted subspace, exhibit a nonlinear dependence on curvature, further validating the modeling approach described by the Frenet-frame equations (\ref{eq:frenet_s}) and (\ref{eq:frenet_eps}).

These results demonstrate that the bilinear formulation effectively captures the state-input coupling, particularly during complex high-speed maneuvers.

\subsection{Cumulative Error Regulator (CER)} \label{sec:CER for trajectory tracking}
Due to the inherent approximation errors associated with representing controlled dynamics through a finite-dimensional Koopman operator, model mismatch occurs. This mismatch inevitably impacts the predictive accuracy of the learned models, resulting in degraded control performance. Tracking algorithms often estimate the position or state of an object based on observations over time. As time passes, minor inaccuracies or errors in these observations or the algorithm's predictions can accumulate. Instead of just tracking the current state, dynamically estimating and compensating for accumulated errors allows for more robust and precise tracking. In vehicle dynamics, such accumulation of errors becomes particularly noticeable in the tracking states of progress along the path ($S$), lateral deviation ($e_y$), and heading error ($e_\psi$), as these states accumulate error over time due to their integrative nature. 
While a common practice to compensate for output tracking error involves penalizing predicted tracking errors at each individual step, introducing an additional block explicitly designed for predicting accumulated error dynamics and constraining this cumulative error to remain near zero at every step allows the control system to detect long-term deviations effectively. Consequently, the system is better able to recognize, respond to, and compensate for model mismatch and prediction inaccuracies.

An augmentation block within the control design, referred to as the CER, was introduced, which is governed by the dynamics
\begin{equation}
    \mathbf{e}_{k+1} = \mathbf{e}_k + \left( \mathbf{y}_k - {\mathbf{y}}^{ref}_k \right),
\end{equation}
where $\mathbf{y}_k = \hat{C} \hat{Z}_{k}$ denotes the predicted tracking outputs (e.g., $\Delta s$, $e_y$, $e_\psi$) , $\hat{C}$ is the matrix mapping $\hat{Z}_k$ to the tracking outputs , and ${\mathbf{y}}^{ref}_k$ is the corresponding output reference trajectory (e.g., ${\Delta s}^{ref}$, ${e_y}^{ref}$, ${e_\psi}^{ref}$). By augmenting the predicted lifted features with the states of error, an extended lifted representation was obtained
\begin{equation}\label{eq:extended_lifted}
    \tilde{Z}_{k} = \begin{bmatrix}
        \hat{Z}_{k}\\
        \mathbf{e}_k
    \end{bmatrix} \in \mathbb{R}^{p + 3},
\end{equation}
which enables the controller to track error accumulation throughout the prediction horizon and penalize deviations from the early stages.
To support this, the input is also augmented to incorporate the desired output reference signal
\begin{equation}
    \tilde{U}_k = \begin{bmatrix}
        U_k \\
        {\mathbf{y}}^{ref}_k
    \end{bmatrix}\in \mathbb{R}^{m+l + 3}.
    \label{eq:augmenetd_input}
\end{equation}
The resulting augmented Koopman system, which integrates the learned Koopman model with the cumulative error module, is formulated as
\begin{align}
\tilde{Z}_{k+1} =
    \tilde{A}\tilde{Z}_{k}+ 
    \tilde{B}\tilde{U}_k &+
    \begin{bmatrix}
        \sum_{i=1}^m H_i \left[  U^i_{k}\cdot \hat{Z}_{k}  \right]\\[4pt]
        0_{3\times 1}
    \end{bmatrix}, \\[6pt]
\tilde{y}_k &=
    \begin{bmatrix}
        \hat{\mathbf{x}}_k\\[4pt]
        e_k
    \end{bmatrix} =
    \tilde{C} \tilde{Z}_{k}.
\end{align}
where
\begin{equation}
\begin{aligned}
    \tilde{A} =&
    \begin{bmatrix}
    A & 0_{3\times3}\\
    \hat{C} & I_{3}
    \end{bmatrix},\ 
   &&\tilde{B}=
    \begin{bmatrix}
    B & 0_{p\times3}\\
    0_{3\times 3} & -I_{3}
    \end{bmatrix} , \\
   \hat{C}=&
   \begin{bmatrix}
       0_{3 \times 3} & I_3 & 0_{3\times (p-6)}\\
   \end{bmatrix},\
   &&\tilde{C}=
    \begin{bmatrix}
    C & 0_{6\times3}\\
    0_{3\times p} & I_{3\times3}
    \end{bmatrix}.
    \end{aligned}
\end{equation}
\subsection{CER-MDBK-MPC} \label{sec:CER-MDBK_MPC}
To evaluate the closed-loop performance of the augmented Koopman model, the \textit{CER MDBK Model Predictive Controller (CER-MDBK-MPC)} framework was introduced.
The integration of the CER with the identified Koopman model forms the basis of the predictive model used within the MPC framework. As illustrated in the control block of Fig.~\ref{fig:network_architecture}, the augmented Koopman system simultaneously propagates both the observable dynamics and the cumulative tracking error at each time step. This joint propagation enables the MPC to account for accumulated deviations from the reference trajectory over the prediction horizon $N$. For a given time $t$, the MPC problem is formulated as a finite-horizon constrained optimal control problem expressed as
\begin{subequations} \label{eq:mpc_bilinear_opt}
% --- Cost function (39a) ---
\begin{align}
\label{eq:mpc_bilinear_opt_cost}
\min_{ \mathbf{u}_t, \dots, \mathbf{u}_{t+N-1}} \quad
& \sum_{k=t}^{t+N-1} \left[ \left( \tilde{y}_{k} - \tilde{r}_{k} \right)^\top Q \left( \tilde{y}_{k} - \tilde{r}_{k}\right) + {\mathbf{u}_{k}}^\top R \mathbf{u}_{k} \right]  \nonumber \\ 
& + \left( \tilde{y}_{t+N} - \tilde{r}_{t+N} \right)^\top Q_N \left( \tilde{y}_{t+N}  - \tilde{r}_{t+N} \right)
\end{align}
% --- Constraints (39b) ---
\begin{align}
\label{eq:mpc_bilinear_opt_constraints}
\text{subject to} \quad 
& \tilde{Z}_{k+1} = \tilde{A} \tilde{Z}_k + \tilde{B} \tilde{U}_k
+ \begin{bmatrix}
\sum_{i=1}^m H_i \left[ U^i_k \cdot \hat{Z}_k \right] \\
0_{3\times 1}
\end{bmatrix}, \nonumber \\
& \tilde{y}_k = \tilde{C} \tilde{Z}_k, \nonumber \\
& \mathbf{u}_{\min} \leq \mathbf{u}_k \leq \mathbf{u}_{\max}, \nonumber \\
& \tilde{Z}_t = 
\begin{bmatrix}
Z_t \\
\mathbf{e}_t
\end{bmatrix}, \quad 
\forall k = t, \dots, t+N-1.
\end{align}
\end{subequations}

The decision variables are the control inputs  \( \mathbf{u}_{k} \), which include steering wheel angle (\( \delta_w\)) and longitudinal driver command (\( u_\zeta \)), to be optimized. To explicitly enforce cumulative tracking errors to approach zero, the reference signal is defined as $\tilde{r}_k=[r_{k}^\top,0_{1\times3}^\top]^\top$ where $r_{k}$ denotes the desired reference trajectory.
At each control step, the measured vehicle states are first mapped into the lifted subspace $Z_t$ based on (\ref{eq: vector of observables}), and the extended lifted representation $\tilde{Z}_t$ is constructed as described in (\ref{eq:extended_lifted}). The optimization problem is then formulated and solved within this augmented lifted space, yielding an optimal control sequence \( \{ \mathbf{u}_k \}_{k=t}^{t+N-1} \) over the prediction horizon, from which only the first control input is applied to the system at each iteration.

It is worth noting that the same optimal control formulation can be employed to develop a CER-MDK-MPC framework by utilizing the trained model based on a linear Koopman structure. This integration of CER with the previously developed MDK-Net model offers a fair comparison baseline for evaluating the impact of bilinear terms in predictive modeling.

Due to the bilinear structure, the dynamics in (\ref{eq:mpc_bilinear_opt_constraints}) result in a non-convex optimization problem. This non-convexity arises from the multiplicative coupling between lifted states and control inputs in the update equation, which increases the computational complexity of MPC. To retain convexity, a common linearization strategy is employed, where the bilinear Koopman dynamics are linearized by fixing the lifted state in bilinear term to its initial value, \( \hat{Z}_{k}|_{t}^{t+N-1} = Z_t \), throughout the prediction horizon, as proposed in \cite{zhao2024deep}. The resulting linear dynamic approximates the behaviors of bilinear realization in the neighborhood of the initial step evaluation. As a result, the equality constraint enforcing the bilinear Koopman dynamic sub-block can be updated as
\begin{align}
\hat{Z}_{k+1} &= A \hat{Z}_{k}+B U_k + \sum_{i=1}^{m} H_i \left[Z_t \, \mathbf{u}^i_{k} \right], \\ \nonumber 
&= A \hat{Z}_k + B_u \mathbf{u}_k+ B_w \mathbf{w}_k+ \sum_{i=1}^{m} \left[ \sum_{j=1}^{p} h_{ij} Z_{t,j} \, \mathbf{u}^{i}_{k} \right], \\ \nonumber 
&= A \hat{Z}_k + B_u \mathbf{u}_k+ B_w \mathbf{w}_k + \sum_{j=1}^{p} \left[ \sum_{i=1}^{m} h_{ij} Z_{t,j}\mathbf{u}^{i}_{k}  \right], \\ \nonumber 
&= A \hat{Z}_k + B_u \mathbf{u}_k+ B_w \mathbf{w}_k + \sum_{j=1}^{p} \hat{H}_j [Z_{t,j}] \mathbf{u}_k, \\ \nonumber 
&= A \hat{Z}_k + B_w \mathbf{w}_k + \left( B_u + \sum_{j=1}^{p} \hat{H}_j [Z_{t,j}] \right) \mathbf{u}_k,
\label{eq:mpc_bilinear_linearized}
\end{align}
where \( H_i= [h_{i1},\dots, h_{ip}]\), \( \hat{H}_j= [h_{1j},\dots, h_{mj}]\), and \( Z_{t,j} \) is the \( j^\text{th} \) component of the lifted features at initial value. The vectors \( \mathbf{w}_k \) denote road exogenous input (curvature here), which is treated as a known disturbance over the prediction horizon. The matrix \( \hat{H}_j[Z_{t,j}] \in \mathbb{R}^{p \times m} \) denotes the aggregated bilinear term evaluated at the initial observable vector evaluation. By incorporating this approximation, the MPC problem becomes convex and can be solved efficiently at each time step. 

%%%%%%%%%%%%%%%%%%%%%%%%%%%%%%%%%%%%%%%%%%%%%%%%%%%%%%%%%%%%%%%%%%
\section{Experimental Setup and Results} \label{sec:Results And 
Experimental Setup}
%%%%%%%%%%%%%%%%%%%%%%%%%%%%%%%%%%%%%%%%%%%%%%%%%%%%%%%%%%%%%%%%%%

The proposed controller is evaluated using a simulated double lane change (DLC) scenario generated in CarSim. A high-fidelity four-wheel vehicle model from CarSim serves as the target plant, with road curvature data obtained via the Advanced Driver Assistance Systems (ADAS) sensor interface. For performance benchmarking, three baseline controllers are considered: (i) the EDMDK-MPC controller based on the approach in \cite{kim2023koopman}, (ii) an LTI-based MPC controller designed using a state-space model identified via MATLAB’s System Identification Toolbox, and (iii) the CER-MDK-MPC controller, which augments the linear Koopman framework from prior work \cite{abtahi2025multi} with the CER block. The latter enables a fair comparison between the proposed CER-MDBK-MPC and its linear counterpart by isolating the effect of bilinear modeling.
All controllers are evaluated under identical conditions with a sampling time of 25 milliseconds, using the same reference trajectory with a constant road curvature profile of \( \kappa = 0.001 \). The prediction horizon $N$ is set to 20 steps. The same optimal control problem formulation is employed for all controllers, with identical weighting matrices.

\begin{figure}
    \centering
    \includegraphics[width=\linewidth]{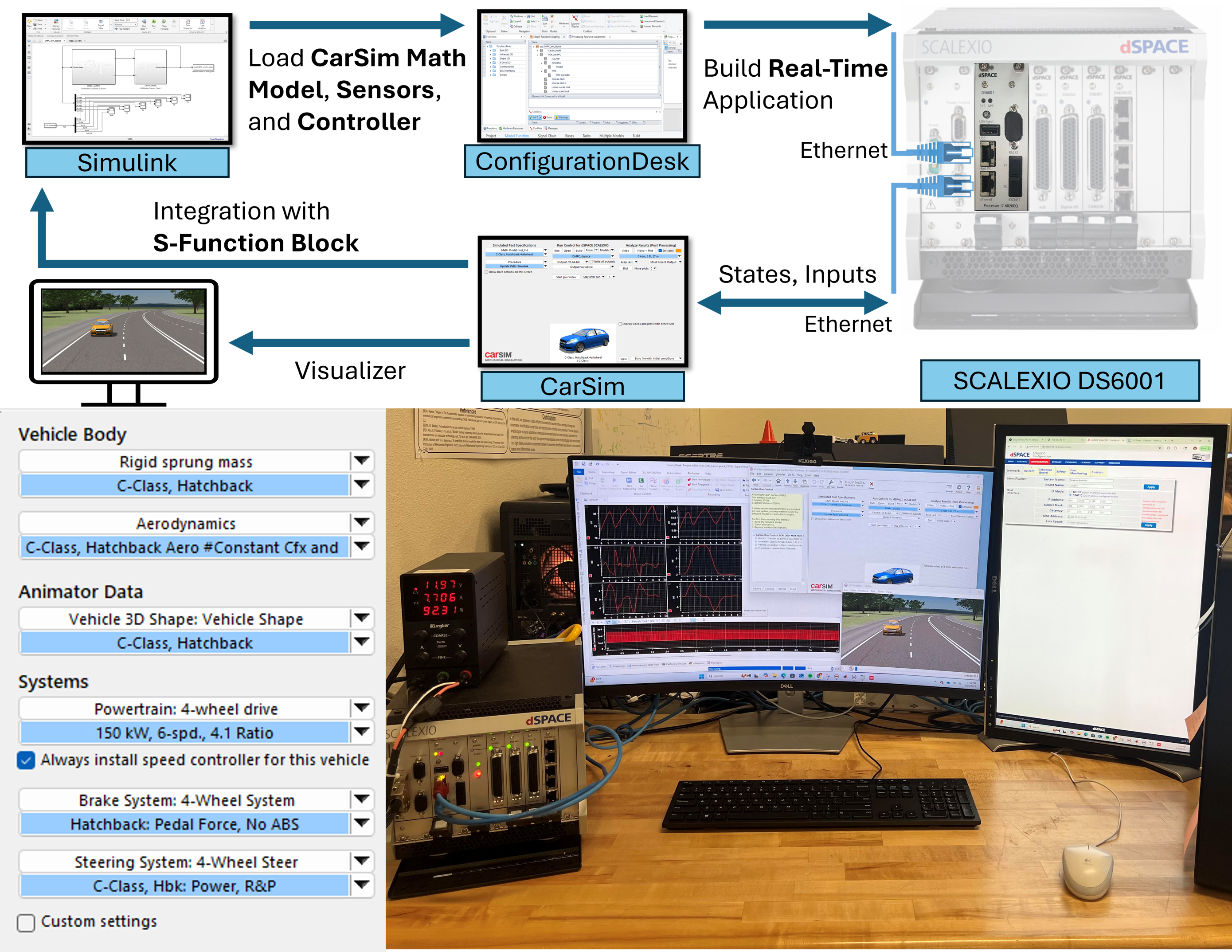}
    \caption{Hardware-in-the-loop (HIL) setup used to evaluate the real-time performance of each controller. The setup integrates Simulink, ConfigurationDesk, and CarSim with the dSPACE SCALEXIO DS6001 real-time target machine. The controllers are tested under realistic vehicle dynamics using high-fidelity simulation and physical I/O connections.}
    \label{fig:HIL_setup}
\end{figure}

All controllers are evaluated on the same dSPACE SCALEXIO DS6001 RT system, which features a 4-core CPU with 8~MB of L3 cache. The system operates in conjunction with CarSim RT software, utilizing the same 25 ms sampling rate, to provide a real-time vehicle model of an all-wheel-drive C-Class hatchback. The hardware-in-the-loop (HIL) configuration is shown in Fig.~\ref{fig:HIL_setup}. A Simulink model integrating both the controller and the plant is deployed onto the SCALEXIO hardware using dSPACE ConfigurationDesk. Communication between the SCALEXIO system and both the CarSim RT software and the host PC is handled via Ethernet. The CarSim visualizer serves as the graphical interface to display real-time vehicle behavior during testing.
\begin{figure}[t]
    \centering
    \includegraphics[width=\linewidth]{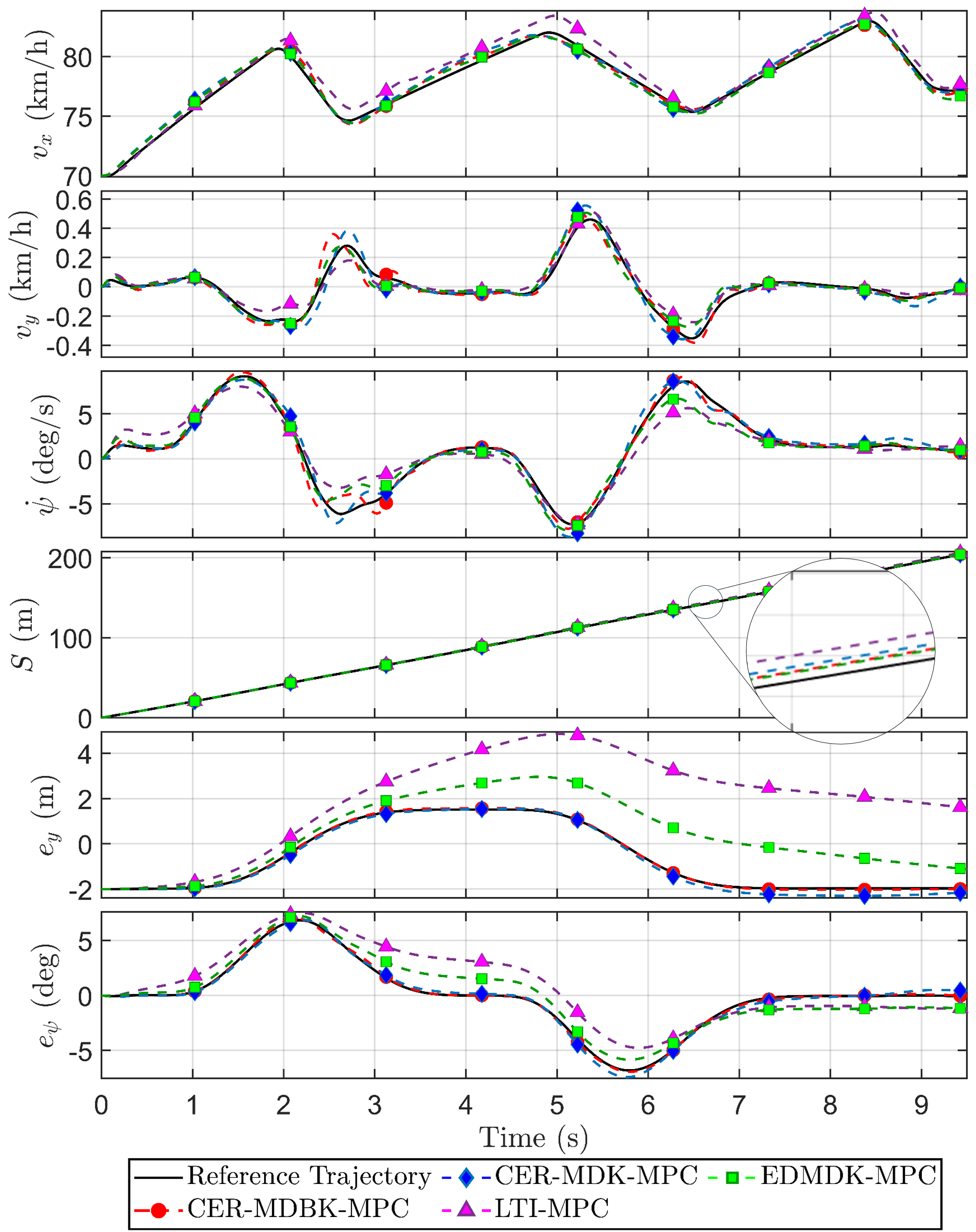}
    \caption{ Comparative performance of CER MDK MPC, EDMDK MPC, CER MDBK MPC, and LTI MPC controllers during a double lane change maneuver. The plots depict the evolution of vehicle states including longitudinal velocity $v_x$, lateral velocity $v_y$, yaw rate $\dot{\psi}$, path progress $S$, lateral error $e_y$, and heading error $e_\psi$ over time.}
    \label{fig:mpc_bilinear_comp}
\end{figure}
Figure~\ref{fig:mpc_bilinear_comp} presents the performance of all four controllers in tracking the reference trajectory, and the RMSE of predictive controllers across all state trajectories is provided in Table~\ref{tab:rmse_controllers}. The LTI-MPC exhibits poor tracking accuracy due to the limited fidelity of the identified LTI model in the original low-dimensional state space, despite being trained on the same dataset as the other methods. However, CER-MDBK-MPC more closely follows the reference trajectory across nearly all tracking states of \(S\), \(e_y\), and \(e_\psi\), with particularly strong performance during the second aggressive lane change maneuver, where other approaches exhibit noticeable deviation. Table~\ref{tab:rmse_controllers} shows that CER-MDBK-MPC significantly reduces the tracking errors of \(S\), \(e_y\), and \(e_\psi\) by 5\%, 97\%, and 88\% compared to the EDMDK-MPC. Meanwhile, these tracking errors were reduced by 35\%, 77\%, and 59\% compared to the previously studied MDK-MPC. The observed improvements highlight the effectiveness of the bilinear terms introduced in the proposed framework, which enhance performance, especially when the vehicle operates under highly nonlinear maneuvers.

For longitudinal velocity \(v_x\), both the figure and RMSE values indicate that the proposed CER-MDBK performs comparably to the baseline EDMDK and CER-MDK algorithms. This observation is consistent with the open-loop prediction results, where the absence of significant bilinear effects in longitudinal dynamics leads to similar performance across all methods. In terms of lateral velocity \(v_y\), CER-MDBK achieves performance similar to that of CER-MDK, while yielding a 5\% improvement in yaw rate tracking \(\dot{\psi}\). Meanwhile, the proposed controller improved the lateral velocity and yaw rate tracking by 15\% and 57\% respectively compared to baseline LTI method.  

The proposed method achieves substantial improvements in tracking accuracy, with up to 97\% error reduction observed in specific state variables compared to the LTI-MPC baseline. To evaluate the overall performance of the proposed CER-MDBK-MPC controller, the average relative reduction in tracking error across all state variables was computed. Compared to the CER-MDK-MPC baseline, the CER-MDBK-MPC achieves a 33\% improvement, while it outperforms the EDMDK and LTI-based controllers by 38\% and 68\%, respectively.

\begin{table}[ht]
\centering
\caption{RMSE of vehicle states with respect to reference trajectory}
\label{tab:rmse_controllers}
\begin{tabular}{lcccc}
\toprule
\textbf{State} & \textbf{CER-MDBK} & \textbf{CER-MDK} & \textbf{LTI} & \textbf{EDMDK} \\ 
\midrule
$v_x$ (km/h) & \textbf{0.332} & 0.425 & 0.821 & 0.333 \\[2pt]
$v_y$ (km/h) & \textbf{0.041} & \textbf{0.041} & 0.048 & 0.043 \\[2pt]
$\dot{\psi}$ (deg/s) & \textbf{0.569} & 0.605 & 1.346 & 0.910 \\[2pt]
$S$ (m) & \textbf{0.206} & 0.318 & 1.186 & 0.211 \\[2pt]
$e_y$ (m) & \textbf{0.037} & 0.165 & 3.105 & 1.236 \\[2pt]
$e_{\psi}$ (deg) & \textbf{0.112} & 0.277 & 1.779 & 0.986 \\ \\
\bottomrule
\end{tabular}
\end{table}
The CER-MDBK-MPC effectively mitigates deviations by capturing nonlinear dynamics through its bilinear Koopman representation. Furthermore, its cumulative error regulator compensates for error propagation during the final stages of the simulation. This capability is clearly illustrated by improved lateral displacement and heading error tracking in the bottom subplots in Fig. \ref{fig:mpc_bilinear_comp}.

\begin{figure}
    \centering
    \includegraphics[width=\linewidth]{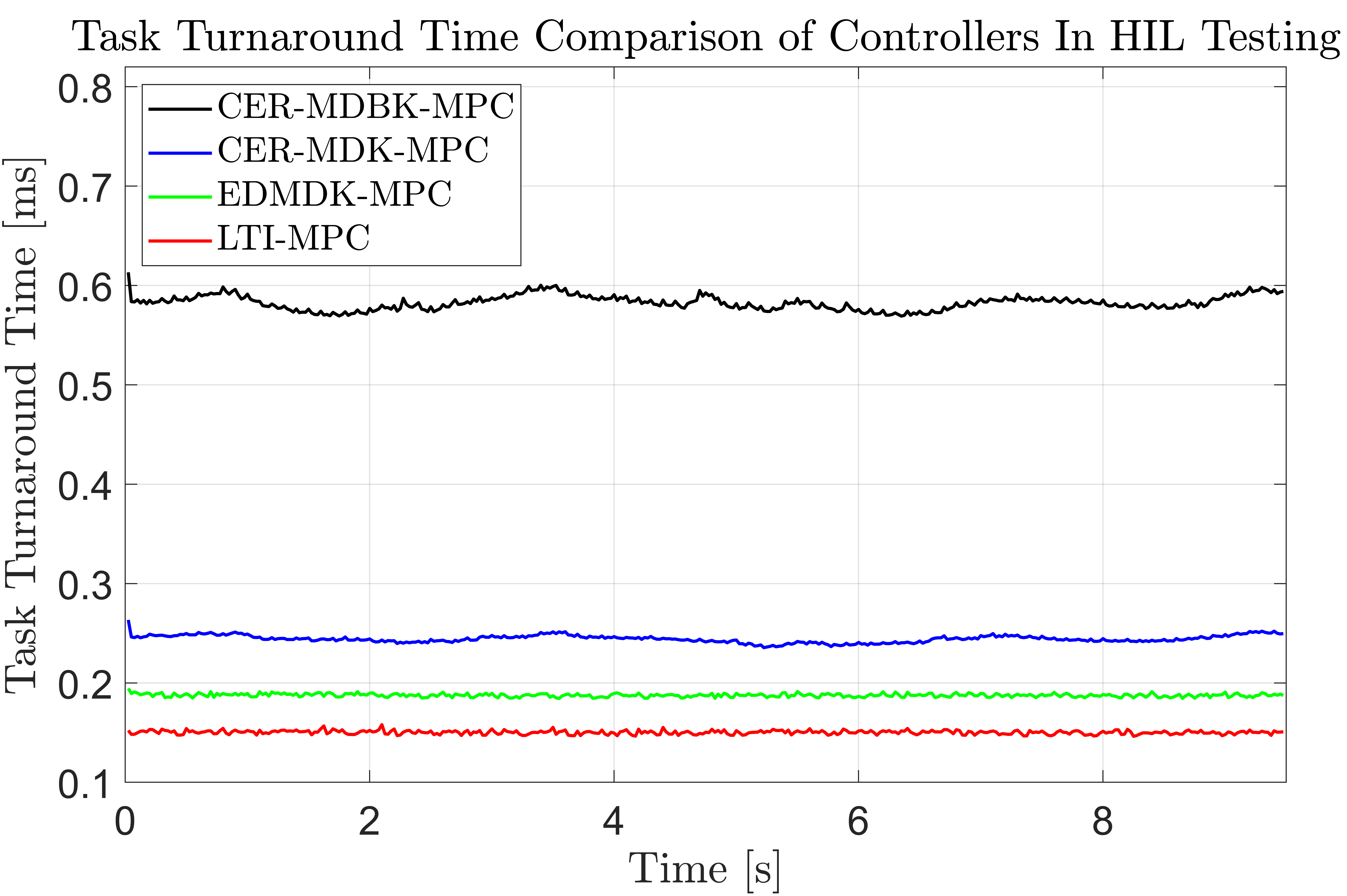}
    \caption{Task turnaround time comparison of all controllers during HIL testing over a 9.5-second simulation. Although CER-MDBK-MPC exhibits the highest computational load (~0.6 ms), all controllers meet real-time requirements well below the 25 ms control step used in the experiment.}
    \label{fig:tTT}
\end{figure}

Figure~\ref {fig:tTT} illustrates the real-time computational performance of the CER-MDBK-MPC controller, as measured by task turnaround time on the SCALEXIO real-time system, compared to baseline controllers operating under a 25 ms sampling interval. The CER-MDBK-MPC achieves an average computation time of 0.58 ms per interval, demonstrating that it runs over 40 times faster than a real-time application. This is followed by CER-MDK-MPC at 0.24 ms, EDMDK-MPC at 0.18 ms, and LTI-MPC at 0.15 ms. The computational effort for CER-MDBK-MPC stems from the bilinear Koopman structure, which necessitates stepwise approximations, unlike the linear predictive structures of CER-MDK and EDMDK. Although LTI-MPC exhibits the lowest computation time due to its simpler, low-dimensional model, CER-MDBK-MPC offers superior tracking accuracy, making it advantageous for real-time, safety-critical applications that require precise modeling of nonlinear dynamics.

\section{Conclusion}\label{sec:Conclusion}
% In this paper, we proposed a Multi-Step Deep Bilinear Koopman Network to learn a globally linear approximation of vehicle dynamics in the curvilinear Frenet frame. The model explicitly captures bilinear interactions between control inputs and lifted state features, which is confirmed by analyzing the learned bilinear interaction matrices. These interactions closely align with the inherent nonlinearities in the vehicle’s dynamic equations, demonstrating the effectiveness of the proposed deep network architecture.
In this paper, a Multi-Step Deep Bilinear Koopman Network was presented that jointly identifies the lifting functions through an encoder structure and learns the corresponding bilinear Koopman matrices. The proposed algorithm explicitly encodes bilinear term coefficients that capture interactions between control inputs and lifted states, closely aligning with the nonlinear cross terms of the inputs in the vehicle’s equations of motion. To further enhance model fidelity, the approach incorporates driver steering, throttle, and brake commands, along with road curvature inputs, to capture nonlinearities in the vehicle drive system and its curvilinear representation. A comprehensive set of loss functions is employed to ensure both short-term and long-term prediction accuracy while promoting model stability. A Cumulative Error Regulator is integrated into the predictive control framework to mitigate prediction errors in tracking dynamics.

The MDBK-Net demonstrated superior prediction accuracy in open-loop evaluations during a single lane change scenario, outperforming baseline approaches including the MDK-Net model and an EDMD-based Koopman model. The closed-loop performance of the CER-MDBK-MPC was evaluated through hardware-in-the-loop testing on a dSPACE SCALEXIO AutoBox RT platform, with a CarSim RT vehicle model used as the target plant. Experimental results showed that the proposed controller improved tracking accuracy by 33\%, 38\%, and 68\% compared to the CER-MDK-MPC, EDMDK-MPC, and LTI-MPC controllers, respectively. Furthermore, the controller maintained real-time feasibility, computing control actions on average 40 times faster than the 25 ms sampling interval.

The proposed approach highlights how deep bilinear structures can efficiently capture the nonlinear characteristics of high-fidelity vehicle dynamics, improving modeling fidelity within predictive control frameworks. By leveraging artificial intelligence to learn expressive yet linear representations of complex dynamics, this work demonstrates a data-driven methodology for achieving more accurate and safer vehicle control. Such improvements in path-following accuracy can directly enhance the safety of autonomous vehicles, particularly during aggressive or high-speed maneuvers.

Future work will focus on extending the bilinear Koopman modeling framework to incorporate real-time sensor noise and disturbances, aiming to improve robustness under practical deployment conditions. Additionally, the application of the Koopman-based approach to more complex maneuvers such as emergency obstacle avoidance and aggressive cornering will be investigated. Finally, integrating domain knowledge through physics-informed encoder structures offers a promising direction for improving model generalization and explicitly capturing critical nonlinearities in the vehicle dynamics.

\ifCLASSOPTIONcaptionsoff
  \newpage
\fi

\bibliographystyle{IEEEtran}
\bibliography{References}

\end{document}